\documentclass[twocolumn,showpacs,preprintnumbers,amsmath,amssymb,superscriptaddress,floatfix]{revtex4}
\usepackage{graphicx}
\usepackage{dcolumn}
\usepackage{bm}
\usepackage{latexsym}
\newcommand{\beq}{\begin{equation}}
\newcommand{\eeq}{\end{equation}}
\newcommand{\ra}{\rightarrow}
\newcommand{\beqn}{\begin{eqnarray}}
\newcommand{\eeqn}{\end{eqnarray}}

\newcommand{\F}{\phantom {1}}
\newcommand{\Kanazawa}{\affiliation{Institute for Theoretical Physics,
Kanazawa University, Kanazawa 920-1192, Japan}}
\newcommand{\RIKEN}{\affiliation{RIKEN Nishina Center for Accelerator-Based Science, Wako, Saitama351-0198, JAPAN}}
\newcommand{\Numazu}{\affiliation{Numazu College of Technology, Numazu 410-8501, Japan}}
\newcommand{\Kochi}{\affiliation{Integrated Information Center, Kochi University, Kochi 780-8520, Japan}}

\begin{document}

\title{Gauge invariance of color confinement due to the  dual Meissner effect\\
caused by Abelian  monopoles}
\author{Tsuneo Suzuki}
\Kanazawa
\RIKEN
\author{Masayasu Hasegawa}
\Kanazawa
\RIKEN
\author{Katsuya Ishiguro}
\Kochi
\RIKEN
\author{Yoshiaki Koma}
\Numazu
\RIKEN
\author{Toru Sekido}
\Kanazawa
\date{\today}

\begin{abstract}
The mechanism of non-Abelian color confinement is studied in SU(2) lattice 
gauge theory in terms of the Abelian fields and monopoles 
extracted from non-Abelian link variables
\textit{without adopting gauge fixing}.
Firstly, the static quark-antiquark potential and force are 
computed with the 
Abelian and monopole Polyakov loop correlators, 
and the resulting string tensions are found to be identical to
the non-Abelian string tension.
These potentials also show the scaling behavior
with respect to the change of lattice spacing.
Secondly, the profile of the color-electric field between a 
quark and an antiquark is investigated with the 
Abelian and monopole Wilson loops.
The color-electric field is squeezed into a flux tube due to
monopole supercurrent with the same Abelian color direction.
The parameters corresponding to the penetration and 
coherence lengths show the scaling behavior, and the 
ratio of these lengths, i.e, the Ginzburg-Landau parameter,
indicates that the vacuum type is near the border of the type~1 and type~2 
(dual) superconductor.
These results are summarized that the Abelian fundamental charge 
defined in an arbitrary color direction is confined 
inside a hadronic state by the dual Meissner effect.
As the color-neutral state in any Abelian color direction 
corresponds to
the physical color-singlet state, 
this effect explains non-Abelian color confinement and 
supports the existence of a gauge-invariant mechanism of
color confinement due to the dual Meissner effect caused by Abelian monopoles. 

\end{abstract}

\maketitle
\section{Introduction}

\par
Color confinement in  quantum chromodynamics (QCD) is still an important 
unsolved  problem~\cite{CMI:2000mp}. 
 't~Hooft~\cite{tHooft:1975pu} 
and Mandelstam~\cite{Mandelstam:1974pi} conjectured that the QCD vacuum 
is a kind of a dual superconducting state caused by condensation 
of magnetic monopoles.
The color charges are then confined inside hadrons due to formation of the 
color-electric flux tube through the dual Meissner effect.
However, in contrast to the Georgi-Glashow 
model~\cite{'tHooft:1974qc,Polyakov:1976fu} 
or SUSY QCD~\cite{Seiberg:1994rs} with scalar fields,
it is not straightforward to identify the color-magnetic monopoles in QCD. 

\par
An interesting idea to realize this conjecture 
is proposed by 't~Hooft~\cite{tHooft:1981ht}, such that
SU(3) QCD can be reduced to an Abelian [U(1)]$^2$ theory 
by adopting a partial gauge fixing, and 
the color-magnetic monopoles appear 
according to $\pi_{2}({\rm SU(3)/[U(1)]^{2}})=\mathbb{Z}^{2}$.
The role of monopoles for the confinement mechanism is investigated
extensively on the lattice by applying 
Abelian projection in the maximally Abelian (MA) 
gauge~\cite{Kronfeld:1987ri,Kronfeld:1987vd},
where monopoles are extracted {\it a la} 
DeGrand-Toussaint~\cite{DeGrand:1980eq}
as in compact U(1) lattice gauge theory.
It is then found that the results strongly
support the dual superconducting 
scenario~\cite{Suzuki:1992rw,Singh:1993jj,Ejiri:1996sh,
Chernodub:1997ay,Bali:1997cp,Suzuki:1998hc,Koma:2003gq,Koma:2003hv}.
The confining properties are dominated by the Abelian 
fields~\cite{Suzuki:1992rw,Kitahara:1994vt,Ejiri:1996sh}
and monopoles~\cite{Stack:1994wm,Shiba:1994ab,Shiba:1994db,Suzuki:1994ay,
Ejiri:1994uw,Ejiri:1996sh}, which are
called Abelian dominance and monopole dominance, respectively.
The color-electric flux is squeezed by the dual Meissner
effect~\cite{Bali:1996mv,Bali:1997cp,Singh:1993jj,Koma:2003gq,Koma:2003hv}.
Moreover monopole condensation is 
confirmed by the energy-entropy balance of the monopole
trajectories~\cite{Shiba:1994db,Kato:1998ur,Chernodub:2000ax,Ishiguro:2001jd}.
These results indicate that 
there must exist a dual Ginzburg-Landau type theory as an infrared
effective theory of QCD~\cite{Ezawa:1982bf,Suzuki:1988yq,Maedan:1988yi}.

\par
However, there are still serious problems to prove this scenario.
Firstly, there are infinite ways of the partial gauge fixing.
Since the behavior of the monopoles can depend on the gauge choice,
it is not clear if the lattice results in the MA gauge are universal.
Note that in the Polyakov (PL) gauge, 
't~Hooft's color-magnetic monopoles~\cite{tHooft:1981ht} 
propagate only in the time direction,
which cannot confine static color charges~\cite{Chernodub:2003mm}.
Secondly, as the 't~Hooft scheme essentially uses the 
Abelian degrees of freedom, it is not explained how non-Abelian color 
charges are confined.

\par
Recently, we have obtained clear numerical evidences of 
Abelian dominance and the dual Meissner effect
in local unitary gauges such as the $F12$ 
and the PL gauges in SU(2) lattice gauge theory~\cite{Sekido:2007mp}, 
where we have used the DeGrand-Toussaint monopoles~\cite{DeGrand:1980eq} 
as in the MA gauge.
These results provide us with the following ideas. 

\begin{enumerate}
\item The DeGrand-Toussaint monopoles on the lattice~\cite{DeGrand:1980eq} 
can be different from the 't~Hooft color-magnetic monopoles~\cite{tHooft:1981ht}.
\item There must exist a gauge-invariant mechanism of color confinement due to Abelian monopoles~\cite{Carmona:2001ja,Cea:2000zr}.
\end{enumerate}

\par
In this paper, we aim to show detailed numerical evidence 
how these ideas are realized.
We investigate the confining properties in SU(2) lattice gauge theory
in terms of the 
gauge-variant
Abelian fields and monopoles extracted from non-Abelian
link variables {\em without adopting any spatially local or non-local 
gauge fixing}.
We find a gauge-invariant Abelian mechanism of color confinement due to Abelian monopoles work even in the continuum limit of SU(2) QCD, although we consider gauge-variant Abelian operators.
The results may also apply to SU(3) gauge theory, since the 
essential features are not altered.

\par
The paper is organized as follows.
In Sec.~\ref{sec:sec2}, we explain how to 
extract the Abelian fields and the monopoles from 
non-Abelian link variables without gauge fixing.
In Sec.~\ref{sec:sec3} and Sec.~\ref{sec:sec4},
we compute the static quark-antiquark potential and the force 
with the Abelian and monopole Polyakov loop correlators, and
find that the string tensions exhibit
Abelian dominance and monopole dominance.
These potentials also show the scaling behavior
with respect to the change of lattice spacing.
In Sec.~\ref{sec:sec5}, we investigate the correlation function 
between the Abelian operators and the Wilson loop.
We observe that the color-electric field is squeezed into a flux tube
due to monopole supercurrent with the same Abelian color direction.
The parameters corresponding to the penetration depth and 
the coherence length show the scaling behavior, and the 
ratio of these lengths, i.e, the Ginzburg-Landau (GL) parameter,
indicates that the vacuum type is near the border of 
the type~1 and type~2 (dual) superconductor.
In Sec.~\ref{sec:sec6}, we discuss implications
of our results, i.e, the Abelian fundamental charge 
defined in an arbitrary color direction is confined 
by the dual Meissner effect.
As the color-neutral state in any Abelian color direction
corresponds to
the physical color-singlet state, 
the dual Meissner effect for the Abelian fundamental charge
can also explain confinement of non-Abelian color charges.
The final section~\ref{sec:sec7} is devoted to conclusion and remarks.
Our preliminary results are already published in Ref.~\cite{Suzuki:2007jp}.

\section{Abelian projection and extraction of monopoles}
\label{sec:sec2}

We explain how to extract the Abelian fields and the color-magnetic
monopoles from the thermalized non-Abelian SU(2) link variables,
\begin{equation}
U_{\mu}(s) = U^0_{\mu}(s)+i\vec{\sigma}\cdot\vec{U}_{\mu}(s) \;,
\end{equation}
where $\vec{\sigma}=(\sigma^{1},\sigma^{2},\sigma^{3})$ 
is the Pauli matrix.
Abelian link variables in one of the color directions, 
for example, in the $\sigma^1$ direction are defined as 
\begin{equation}
u_{\mu}(s) = \cos \theta_{\mu}(s) +i\sigma^1\sin \theta_{\mu}(s) \; ,
\end{equation}
where
\begin{equation}
\theta_{\mu}(s) 
= \arctan\bigl(\frac{U^1_{\mu}(s)}{U^0_{\mu}(s)}\bigr) \; 
\end{equation}
correspond to the Abelian fields.
Without gauge fixing the Abelian fields in any color directions
should be equivalent.

\par
We then define the Abelian field strength tensors as 
\begin{eqnarray}
\Theta_{\mu \nu}(s) 
&=&
\theta_{\mu}(s) + \theta_{\nu}(s+\hat{\mu}) 
-\theta_{\mu}(s+\hat{\nu}) - \theta_{\nu}(s) \nonumber\\ 
&=& \bar{\Theta}_{\mu \nu}(s) + 2\pi n_{\mu \nu}(s) \; ,
\label{abelian-field} 
\end{eqnarray}
where $\bar{\Theta}_{\mu \nu}\in [-\pi, \pi]$ and
$n_{\mu \nu}(s)$ is an integer corresponding to the number of
the Dirac strings piercing the plaquette.
The monopole currents  are then defined by~\cite{DeGrand:1980eq} 
\begin{eqnarray}
k_{\nu}(s) &=& 
{1\over4\pi}\epsilon_{\mu\nu\rho\sigma}
\partial_{\mu} \bar{\Theta}_{\rho \sigma}(s+\hat{\nu}) \nonumber \\
&=&
-{1\over2}\epsilon_{\mu\nu\rho\sigma}
\partial_{\mu} n_{\rho \sigma}(s+\hat{\nu})  \; \in \; \mathbb{Z}\;,
\label{mon-current} 
\end{eqnarray}
where $\partial_{\mu}$ is regarded as a forward difference. 

\section{Abelian dominance}
\label{sec:sec3}

We show the result of the
Abelian static potential~\cite{Suzuki:2007jp}. 
We generate thermalized gauge configurations using the SU(2) Wilson action 
at a coupling constant~$\beta=2.5$ on the lattice~$N_{s}^3 \times N_{t}=24^3 \times 24$, 
where the lattice spacing ~$a(\beta=2.5)=0.0836(8)$~[fm]
is fixed by assuming $\sqrt{\sigma}=440$~[MeV].

\par
By using the multi-level noise reduction 
method~\cite{Luscher:2001up}, we
evaluate the Abelian static potential $V_{\rm A}$ from the 
correlation function of the 
Abelian Polyakov loop operator
\begin{equation}
P_{\rm A} = \exp[i\sum_{k=0}^{N_{t}-1}\theta_4(s+k\hat{4})] \;,
\label{eq-PA}
\end{equation}
separated at a distance $R$ as
\begin{equation}
V_{\rm A}(R_{I}) = -\frac{1}{a N_{t}}\ln \langle  P_{A}(0) P_{A}^{*}(R)\rangle \;.
\end{equation}
The $q$-$\bar{q}$ distance $R$ is improved to
$R_{I}=(4\pi G(R))^{-1}$ in order to reduce the lattice artifact due to finite-lattice spacing,  where $G(R)$ is the Green function of the 
lattice Laplacian in three dimensions~\cite{Necco:2001xg,Luscher:2002qv}.
For the multi-level method, the number of sub-lattices adopted is 6
and the sublattice size is $4a$.

\par
\begin{figure}[t]
\includegraphics[height=5.5cm]{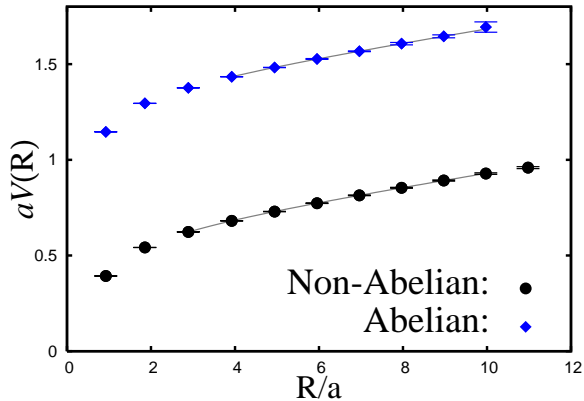}
\caption{The Abelian static potential 
in comparison with the non-Abelian one.
The lines denote the best fitting curve to the function $V_{\rm fit}(R)$.}
\label{fig-1}
\end{figure}

The result is plotted in Fig.~\ref{fig-1}, where the non-Abelian static potential computed from the ordinary
Polyakov loop correlation function is also plotted for comparison.
The number of independent gauge configurations 
is $N_{\rm conf}=10$ in both cases, but the number 
of internal updates in the multi-level method is 
15000 for the non-Abelian case and 160000 for the Abelian case.
The statistical errors are determined by the jackknife method.

\par
We fit the potential to the usual functional form
\begin{eqnarray}
V_{\rm fit}(R)=\sigma R - c/R + \mu \;, 
\label{pot-fit}
\end{eqnarray}
where $\sigma$ denotes the string tension, $c$ the Coulombic coefficient, 
and $\mu$ the constant.
The result is summarized in Table~\ref{stringtension}.
We find Abelian dominance such that the Abelian string tension 
is the same as the non-Abelian one.

Here we make a comment on the theoretical observations of the Abelian dominance of the string tension using the character expansion~\cite{Ogilvie:1998wu,Faber:1998en}. The authors of \cite{Faber:1998en} say that they have proved exactly the Abelian dominance without gauge-fixing based on a relation for any two half-integer representations $j_1>j_2$ 
\beq
       \lim_{T\ra \infty} {W_{j_1}[R,T] \over W_{j_2}[R,T] } = 0,
\label{lim}
\eeq 
where 
\beq
      W_j[C] = {1\over 2j+1} <\chi_j[U(C)]>
\eeq
and $\chi_j[g]$ is the SU(2) group character in representation $j$. In Ref.~\cite{Faber:1998en}, the above relation is derived  on some considerations of screening effects and gluelump energy. But nobody knows an exact method of analytic calculations in the infrared region of QCD, so that their considerations about the screening effects and gluelump energy are not exact theoretically, although very plausible.
Our numerical observations here are hence non-trivial and they suggests the above relation (\ref{lim}) is actually exact. 


\begin{table}[t]
\begin{center}
\caption{\label{stringtension}
Best fitted values of the string tension $\sigma a^2$,
the Coulombic coefficient $c$ and the constant $\mu a$.
$V_{\rm NA}$ denotes the non-Abelian static potential.  
$N_{\rm iup}$ is the number of internal updates in 
the multi-level method. 
FR means the fitting range.
The $\chi ^2$ for the central value 
is $\chi^2/N_{\rm df} < 0.1$.}
\begin{tabular}{l|c|c|c|c|c}
\hline
& $\sigma a^2$ & $c$ & $\mu a$ & FR(R/a) & $N_{\rm iup}$\\ 
\hline
$V_{\rm NA}$ & 0.0348(7)\F  & 0.243(6)\F & 0.607(4)\F & 3.92 - 9.97 & 15000\\
$V_{\rm A}$ & 0.0352(16) & 0.231(39) & 1.357(17) & 4.94 - 9.97 & 160000\\
\hline
\end{tabular}
\end{center}
\end{table}

\section{Monopole dominance}
\label{sec:sec4}

\subsection{The monopole Polyakov loop}

\par
We investigate the monopole contribution to the static potential
in order to examine the role of monopoles for confinement.
The monopole part of the Polyakov loop operator is extracted as follows.
Using the lattice Coulomb propagator $D(s-s')$, which satisfies
$\partial_{\nu}\partial'_{\nu}D(s-s') = -\delta_{ss'}$ with a
forward (backward) difference $\partial_{\nu}$ ($\partial'_{\nu}$), 
the temporal component of the Abelian fields $\theta_{4}(s)$ are written as 
\begin{equation}
\theta_4 (s) 
= -\sum_{s'} D(s-s')[\partial'_{\nu}\Theta_{\nu 4}(s')+
\partial_4 (\partial'_{\nu}\theta_{\nu}(s'))] \; . 
\label{t4}
\end{equation} 
Inserting Eq.~\eqref{t4} (and then Eq.~\eqref{abelian-field})
to the Abelian Polyakov loop~\eqref{eq-PA},
we obtain
\begin{eqnarray}
&&P_{\rm A} = P_{\rm ph} \cdot P_{\rm mon}\; ,\nonumber\\
&&P_{\rm ph} = \exp\{-i\sum_{k=0}^{N_{t}-1} \!\sum_{s'}
D(s+k\hat4-s')\partial'_{\nu}\bar{\Theta}_{\nu 4}(s')\} \; ,\nonumber\\
&&P_{\rm mon} = \exp\{-2\pi i\sum_{k=0}^{N_{t}-1}\! \sum_{s'}
D(s+k\hat4-s')\partial'_{\nu}n_{\nu 4}(s')\}\; .\nonumber\\
\label{ph-mon}
\end{eqnarray}
We call $P_{\rm ph}$ the photon
and $P_{\rm mon}$ the monopole parts of 
the Abelian Polyakov loop, respectively~\cite{Suzuki:1994ay}.
The latter is due to the fact that the Dirac strings 
$n_{\nu 4}(s)$ lead to the monopole currents in 
Eq.~\eqref{mon-current}~\cite{DeGrand:1980eq}.
Note that the second term of Eq.~\eqref{t4} does
not contribute to the Abelian Polyakov loop 
in Eq.~\eqref{eq-PA}.

\begin{table}[t]
\begin{center}
\caption{\label{data}
Simulation parameters for the measurement of the static potential
and the force from $P_{\rm A}$, $P_{\rm ph}$ and $P_{\rm mon}$.
$N_{\rm RGT}$ is the number of random gauge transformations.}
\begin{tabular}{c|c|c|c|c}
\hline
$\beta$ &$N_{s}^{3}\times N_{t}$& $a(\beta)$~[fm]& $N_{\rm conf}$ & $N_{\rm RGT}$ \\ 
\hline
2.20 & $24^{3} \times 4$ & 0.211(7)\F & 6000 & 1000\\
2.35 &$24^{3} \times 6$ & 0.137(9)\F & 4000 & 2000\\
2.35 &$36^{3} \times 6$ & 0.137(9)\F & 5000 & 1000\\
2.43 &$24^{3} \times 8$ & 0.1029(4)& 7000 & 4000 \\
\hline
\end{tabular}
\end{center}
\end{table}

\begin{figure}[t]
\begin{center} 
\includegraphics[height=5.5cm]{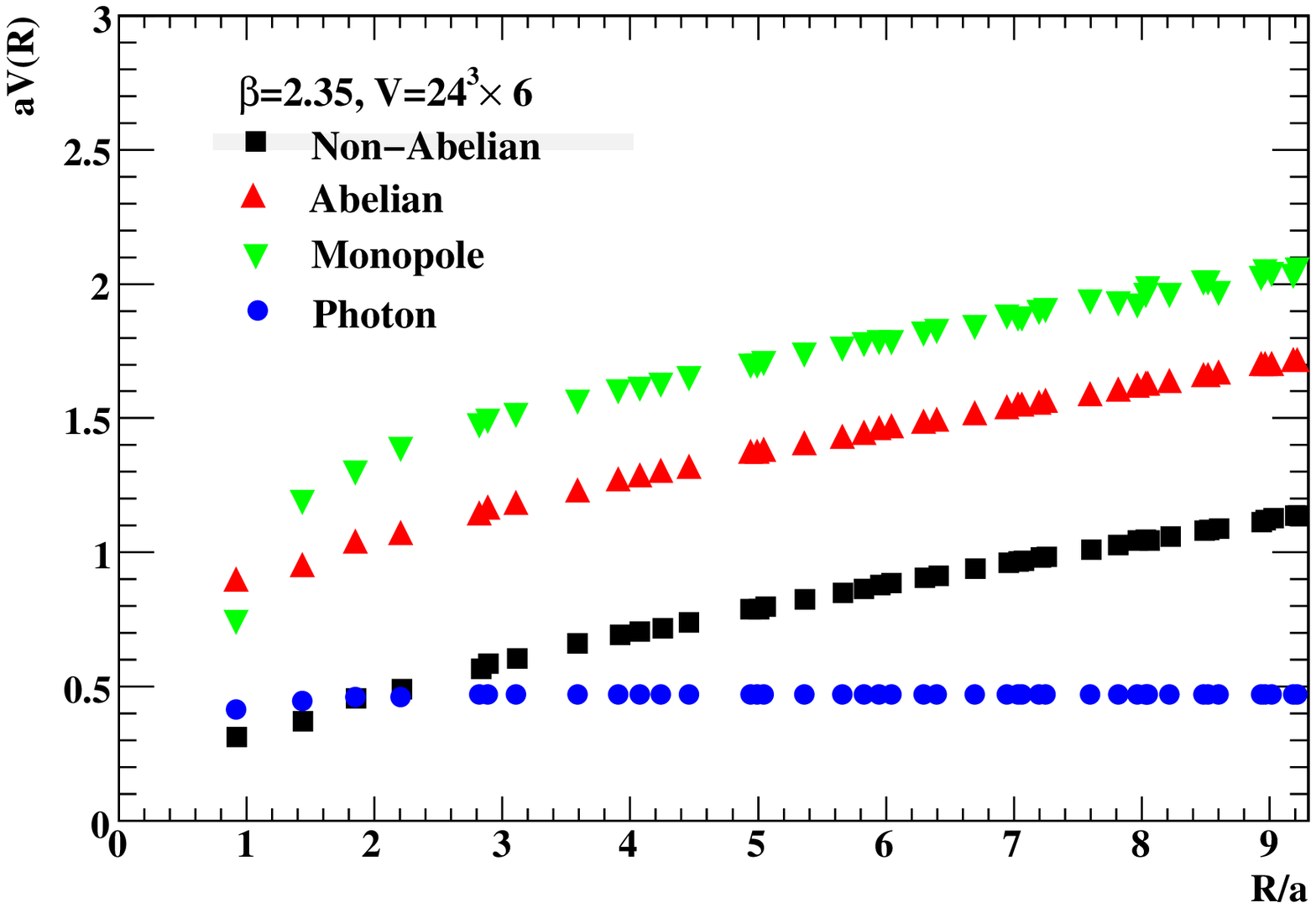}
\includegraphics[height=5.5cm]{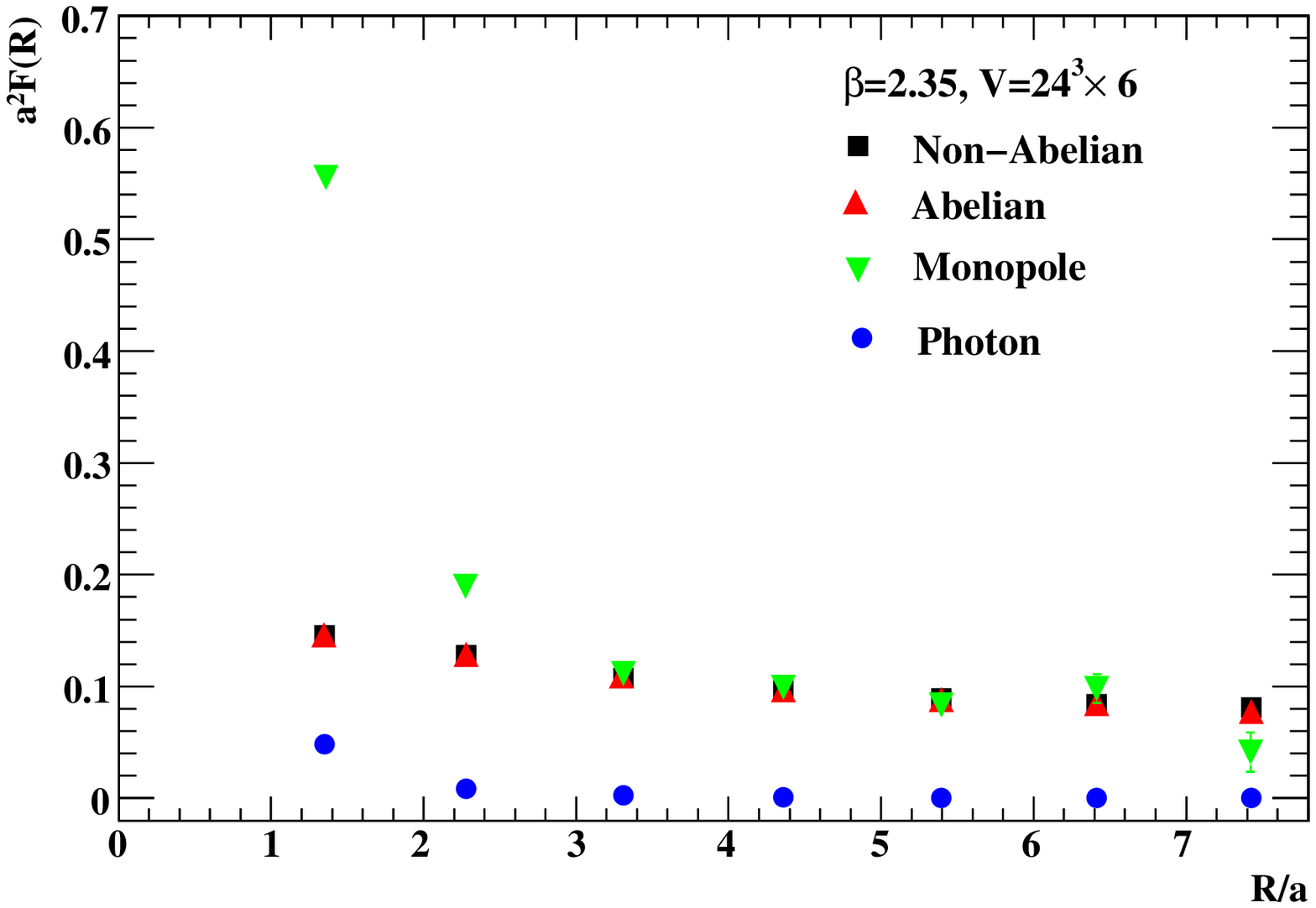}
\caption{\label{fig-2} 
The static potential (top) and the force (bottom)
from the non-Abelian, the Abelian, the monopole and the photon 
Polyakov loop correlation function
at $\beta=2.35$ on the $24^3\times 6$ lattice.}
\end{center}
\end{figure}

\begin{figure}[t]
\begin{center}  
\includegraphics[height=5.5cm]{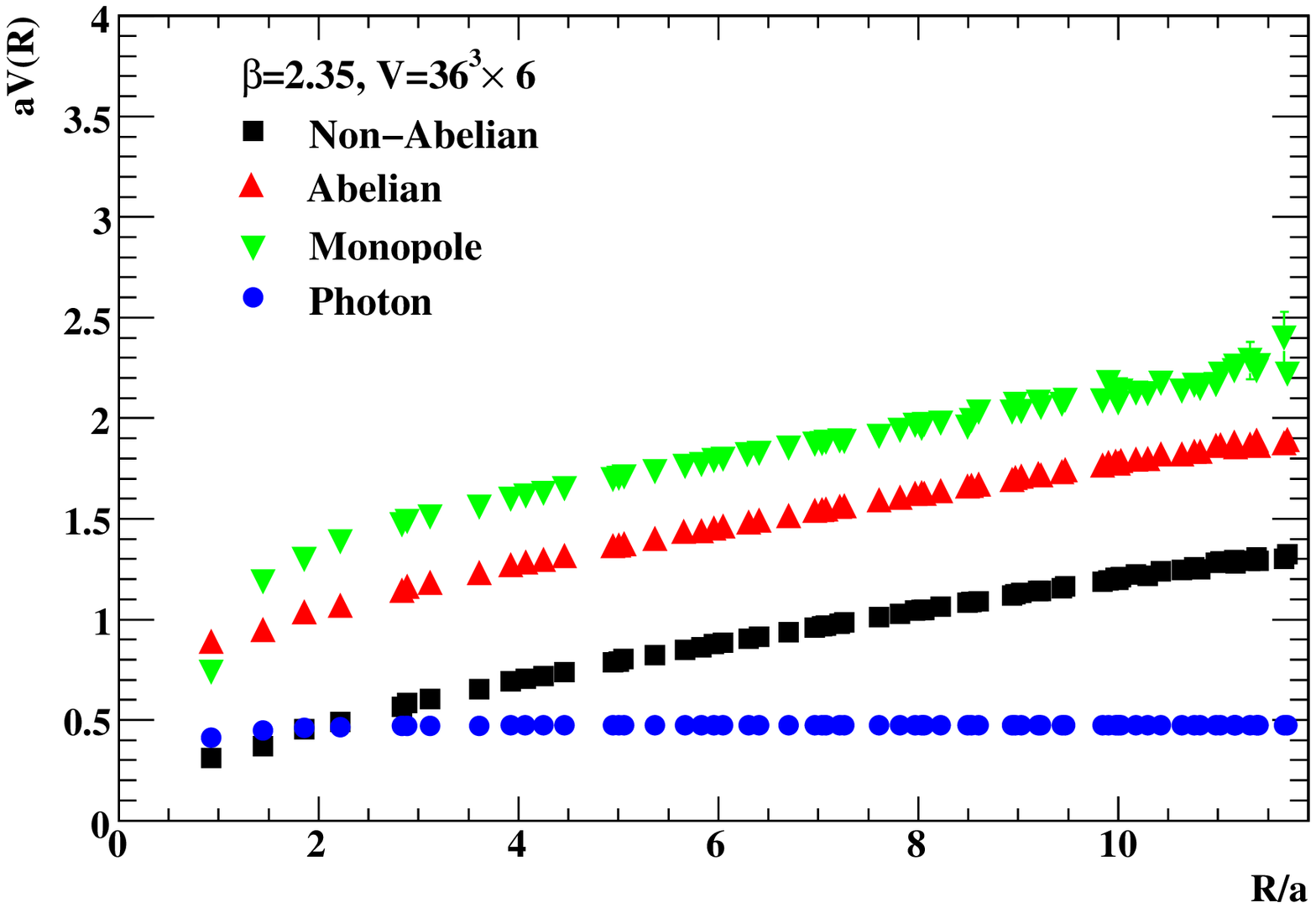}
\includegraphics[height=5.5cm]{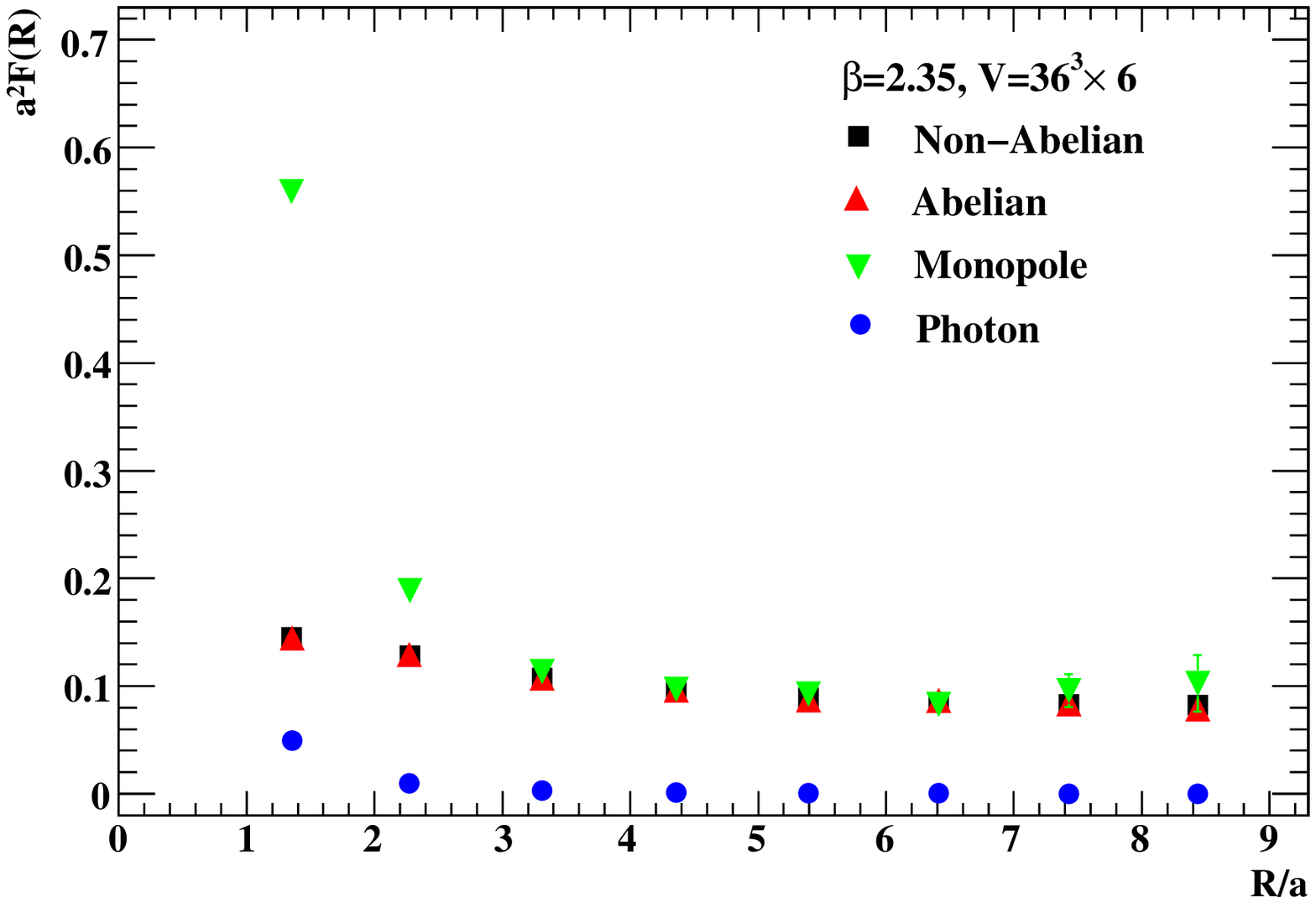}
\caption{\label{fig-5}
The same plot as in Fig.~\ref{fig-2} at
$\beta=2.35$ on the $36^3\times 6$ lattice.}
\end{center}
\end{figure}

\begin{figure}[t]
\begin{center}  
\includegraphics[height=5.5cm]{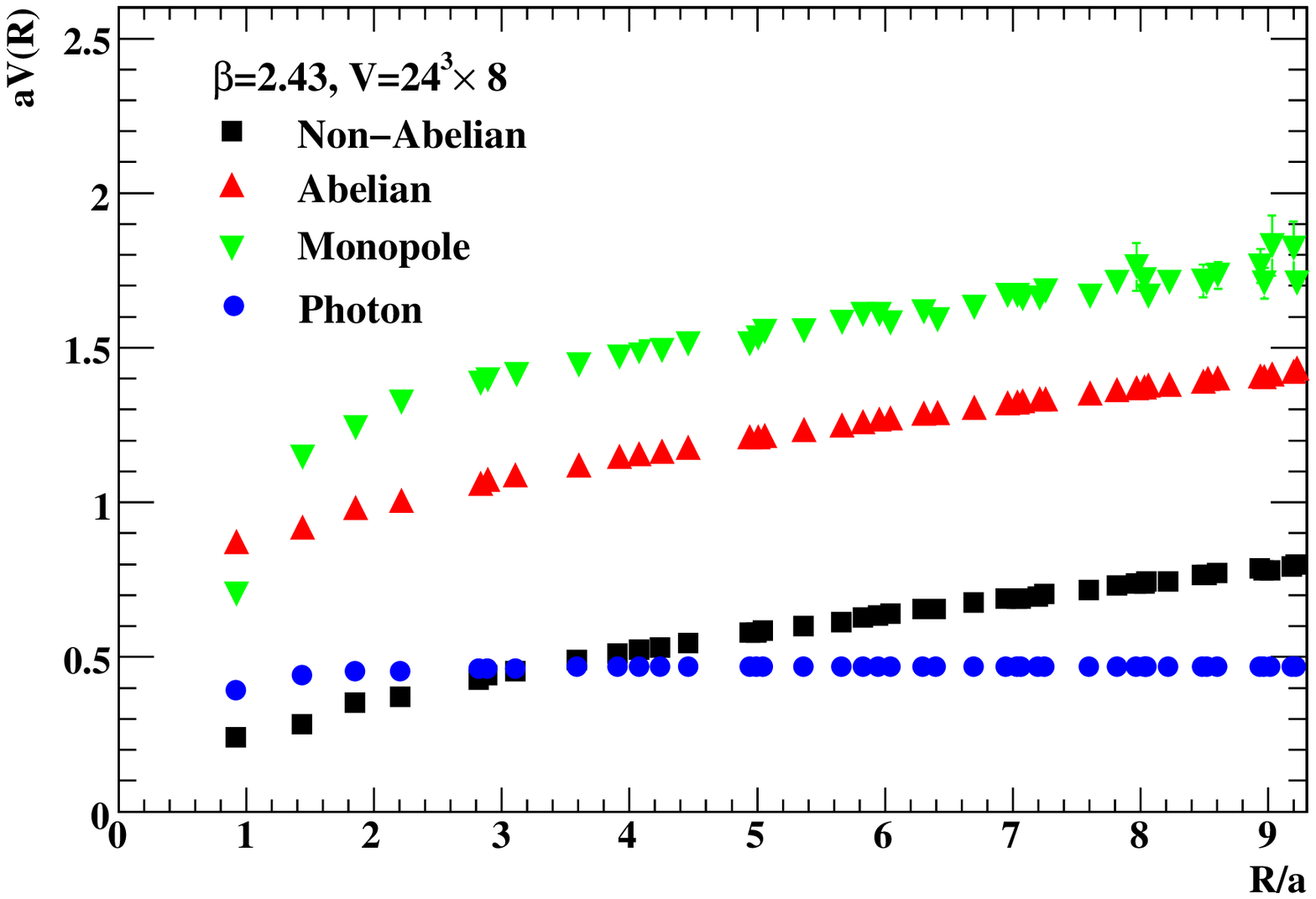}
\includegraphics[height=5.5cm]{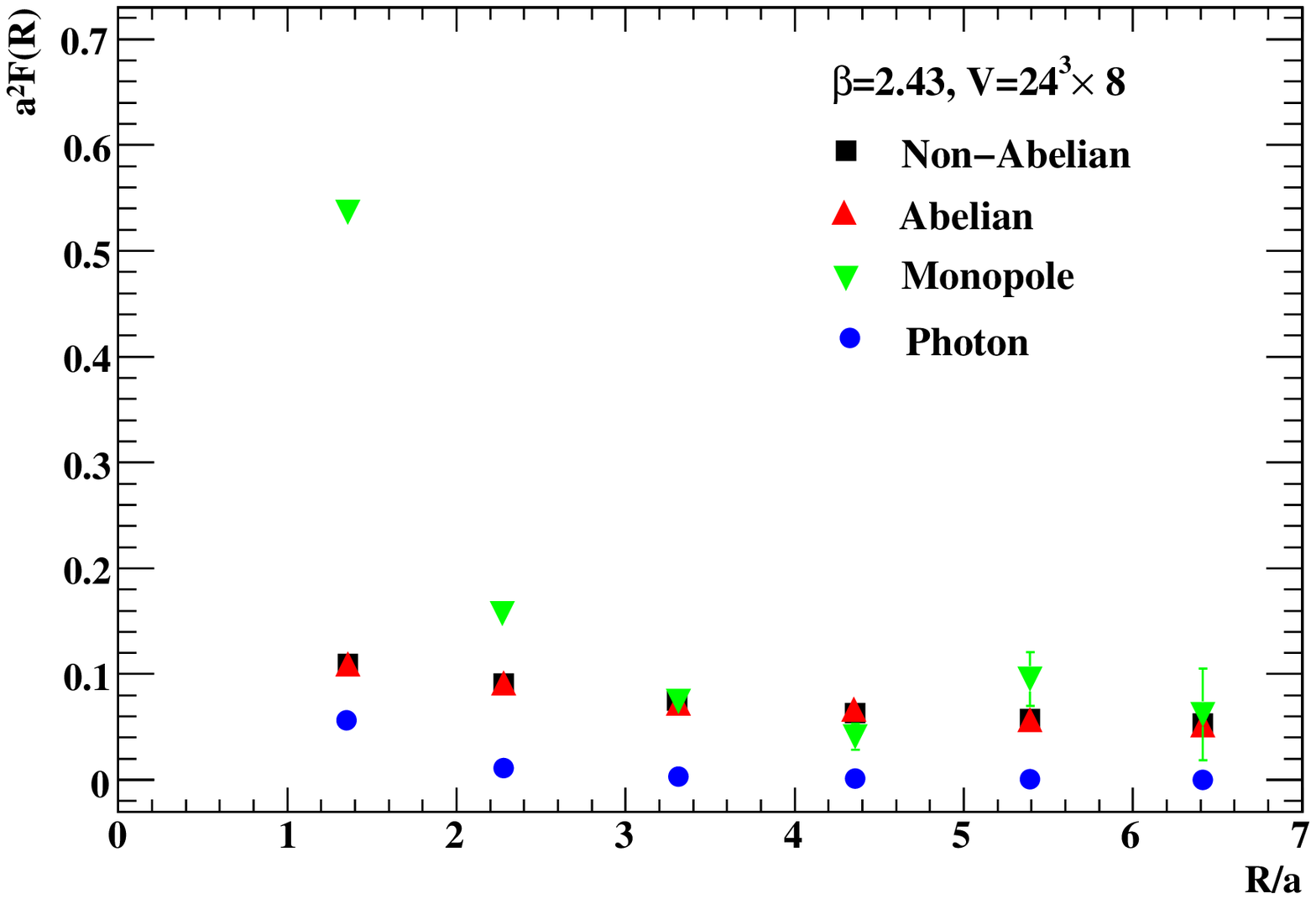}
\caption{\label{fig-3}
The same plot as in Fig.~\ref{fig-2} at
$\beta=2.43$ on the $24^3\times 8$ lattice.}
\end{center}
\end{figure}

\subsection{Simulation parameters}

\par
We then compute the static potential from the monopole 
Polyakov loop correlation function.
However, since Eq.~\eqref{ph-mon} contains
the non-local Coulomb propagator $D(s-s')$
and the Polyakov loop is not written as 
a product of local operators along the time direction,
the multi-level method cannot be applied.
Without such a powerful noise reduction method, 
it is hard to measure the Polyakov loop correlation function
at zero temperature with the present available computer resource.
Thus we consider a finite temperature $T \neq 0$ system 
in the confinement phase.
We set $T = 0.8~T_{c}$.
In order to examine the scaling behavior of the potential,
we simulate the Wilson action on the $24^3 \times (N_t=4,6,8)$ 
lattices.
We choose the gauge coupling for each $N_t$ so as 
to keep the same temperature.
We also investigate the spatial volume dependence of the potential
for the $N_{t}=6$ case.
Simulation parameters are summarized in Table~\ref{data}.
The lattice spacing $a(\beta)$ is determined by using the Sommer 
scale ($r_{0}=0.5$~[fm])  at zero temperature.

\subsection{Noise reduction by gauge averaging}

\par
Since the signal-to-noise ratio of the correlation functions of
$P_{\rm A}$, $P_{\rm ph}$ and $P_{\rm mon}$ are still very small with 
no gauge fixing,
we adopt a new noise reduction method~\cite{Suzuki:2007jp}.
For a thermalized gauge configuration,
we produce many gauge copies applying random gauge transformations.
Then we compute the operator for each copy, and 
take the average over all copies.
It should be noted that as long as a gauge-invariant operator is evaluated, 
such copies are identical, but they are not if a gauge-variant 
operator is evaluated as in the present case.
The results obtained with this method 
are gauge-averaged, thus, gauge-invariant.

\par
In practice, we prepare a few thousand of gauge 
copies for each independent gauge configuration (see Table~\ref{data}).
We also apply one-step hypercubic
blocking (HYP)~\cite{Hasenfratz:2001hp}
to the temporal links for further noise reduction.
The short-distance part of the potential may be affected by HYP.

\subsection{Results}

\par
We obtain very good signals for the potentials and the forces
defined by differentiating the potential with respect to $R$. 
The results at $\beta=2.35$ on the $24^{3} \times 6$ lattice 
and on the $36^{3} \times 6$ lattice,
and at $\beta=2.43$ on the $24^{3} \times 8$ lattice 
are plotted in Figs.~\ref{fig-2},~\ref{fig-5} and~\ref{fig-3}, respectively.
The $q$-$\bar{q}$ distances $R$ of the potentials and the forces
are improved to $R_{I}$ and $\bar{R}=\left(\frac{4\pi}{a}\{G(R-a)-G(R)\}\right)^{-\frac{1}{2}}$, respectively.
We fit these potentials to the function $V_{\rm fit}(R)$ in Eq.~\eqref{pot-fit} 
and extract the string tension and the Coulombic coefficient,
which are summarized in Table~\ref{stringtension_p}.
Since the potential and the force at $\beta=2.20$ on 
the $24^{3} \times 4$ lattice are already published in Ref.~\cite{Suzuki:2007jp},
 we only present the fitting result for this data set.

\begin{table}[t]
\begin{center}
\caption{\label{stringtension_p}
Best fitted values of the string tension $\sigma a^2$, the
Coulombic coefficient $c$, and the constant $\mu a$ for the
potentials $V_{\rm NA}$, $V_{\rm A}$, $V_{\rm mon}$ and $V_{\rm ph}$.}
\begin{tabular}{l|c|c|c|c|c}
\multicolumn{6}{l}{}\\ \hline
$24^3\times 4$& $\sigma a^2$ & $c$ & $\mu a$ & FR($R/a$) 
& $\chi^2/N_{\rm df}$ \\ \hline
$V_{\rm NA}$   & 0.181(8)  & 0.25(15) & 0.54(7)  & 3.9 - 8.5 & 1.00 \\ 
$V_{\rm A}$ & 0.183(8) & 0.20(15) & 0.98(7)  & 3.9 - 8.2 & 1.00 \\ 
$V_{\rm mon}$ & 0.183(6)  & 0.25(11) & 1.31(5)  & 3.9 - 6.7 & 0.98 \\ 
$V_{\rm ph}$ &$-2(1)\times 10^{-4}$ & 0.010(1) & 0.48(1)  & 4.9 - 9.4 & 1.02 \\ \hline
\multicolumn{6}{l}{$24^3\times 6$}\\ \hline
$V_{\rm NA}$  & 0.072(3)  & 0.49(6)\F & 0.53(3)  & 4.0 - 9.0\F & 0.99 \\ 
$V_{\rm A}$ & 0.073(4)  & 0.41(7)\F & 1.09(3)  & 3.7 - 10.9 & 1.00 \\ 
$V_{\rm mon}$ & 0.073(4)  & 0.44(10) & 1.41(4)  & 3.9 - 9.3\F & 1.00 \\ 
$V_{\rm ph}$ &$-1.7(3)\times 10^{-4}$ & 0.0131(1) & 0.4717(3)  & 5.1 - 9.4\F & 0.99\\ \hline
\multicolumn{6}{l}{$36^3\times 6$}\\\hline
$V_{\rm NA}$  & 0.072(3)  & 0.48(9) & 0.53(3)  & 4.6 - 12.1 & 1.03 \\ 
$V_{\rm A}$ & 0.073(2)  & 0.47(6) & 1.10(2)  & 4.3 - 11.2 & 1.03 \\ 
$V_{\rm mon}$ & 0.073(3)  & 0.46(7) & 1.43(3)  & 4.0 - 11.8 & 1.01 \\ 
$V_{\rm ph}$ &$-1.0(1)\times 10^{-4}$ & 0.0132(1) & 0.4770(2)  & 6.4 - 11.5 & 1.03\\
\hline
\multicolumn{6}{l}{$24^3\times 8$}\\ \hline
$V_{\rm NA}$  & 0.0415(9)  & 0.47(2) & 0.46(8)  & 4.1 - 7.8\F & 0.99 \\ 
$V_{\rm A}$ & 0.041(2)  & 0.47(6) & 1.10(3)  & 4.5 - 8.5\F & 1.00 \\ 
$V_{\rm mon}$ & 0.043(3)  & 0.37(4) & 1.39(2)  & 2.1 - 7.5\F & 0.99 \\ 
$V_{\rm ph}$ &$-6.0(3)\times 10^{-5}$ &0.0059(3) & 0.46649(6)  & 7.7 - 11.5 & 1.02\\
\hline
\end{tabular}
\end{center}
\end{table}

\begin{figure}[t]
\begin{center}  
\includegraphics[height=5.5cm]{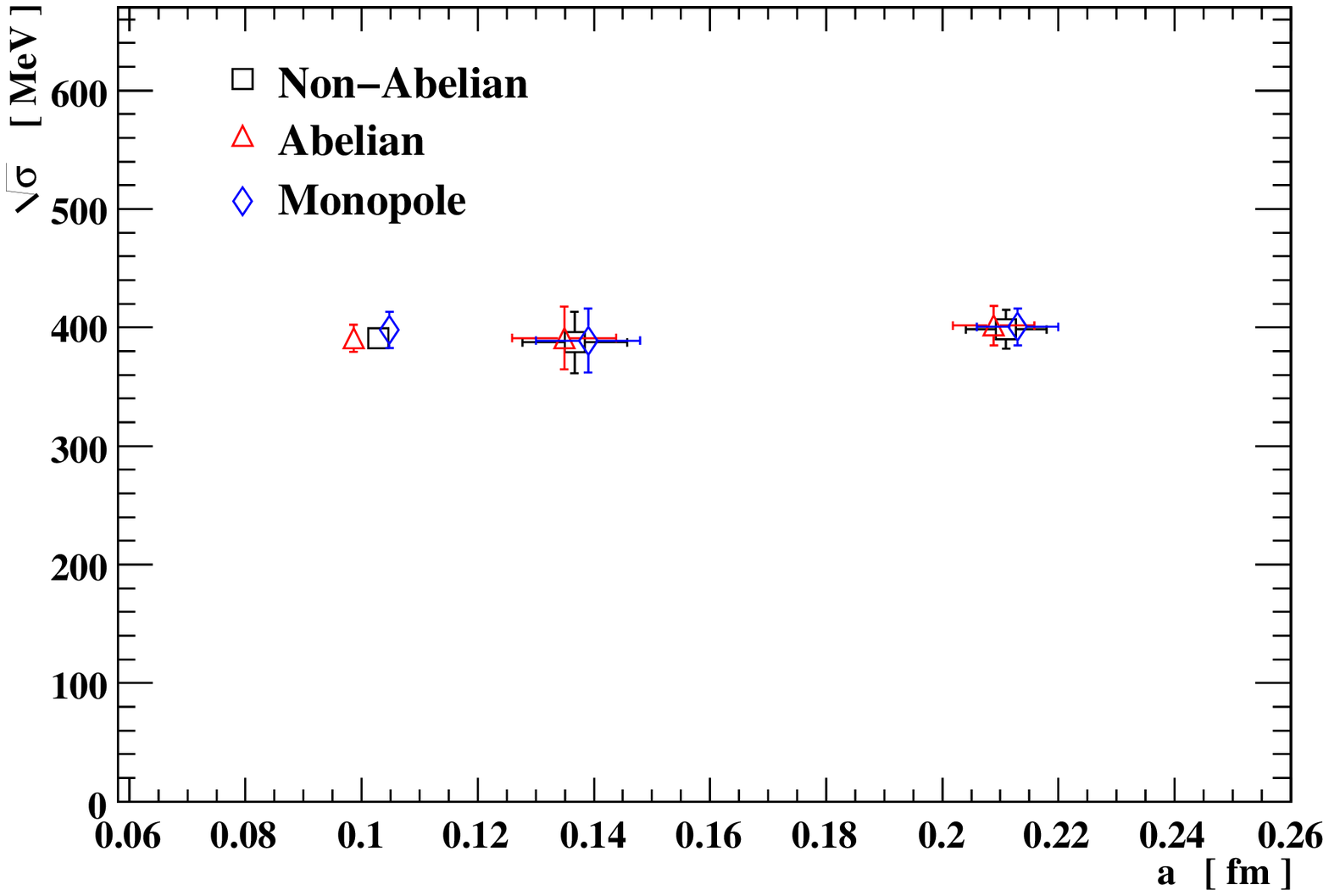}
\caption{\label{fig-4} The $a(\beta)$ dependence 
of the square root of the non-Abelian, Abelian and monopole string tensions
for the same temperature $T=0.8 T_{c}$.
The bottom axis for a set of three data points 
at the same lattice spacing is slightly shifted to distinguish each other.}
\end{center}
\begin{center}  
\includegraphics[height=5.5cm]{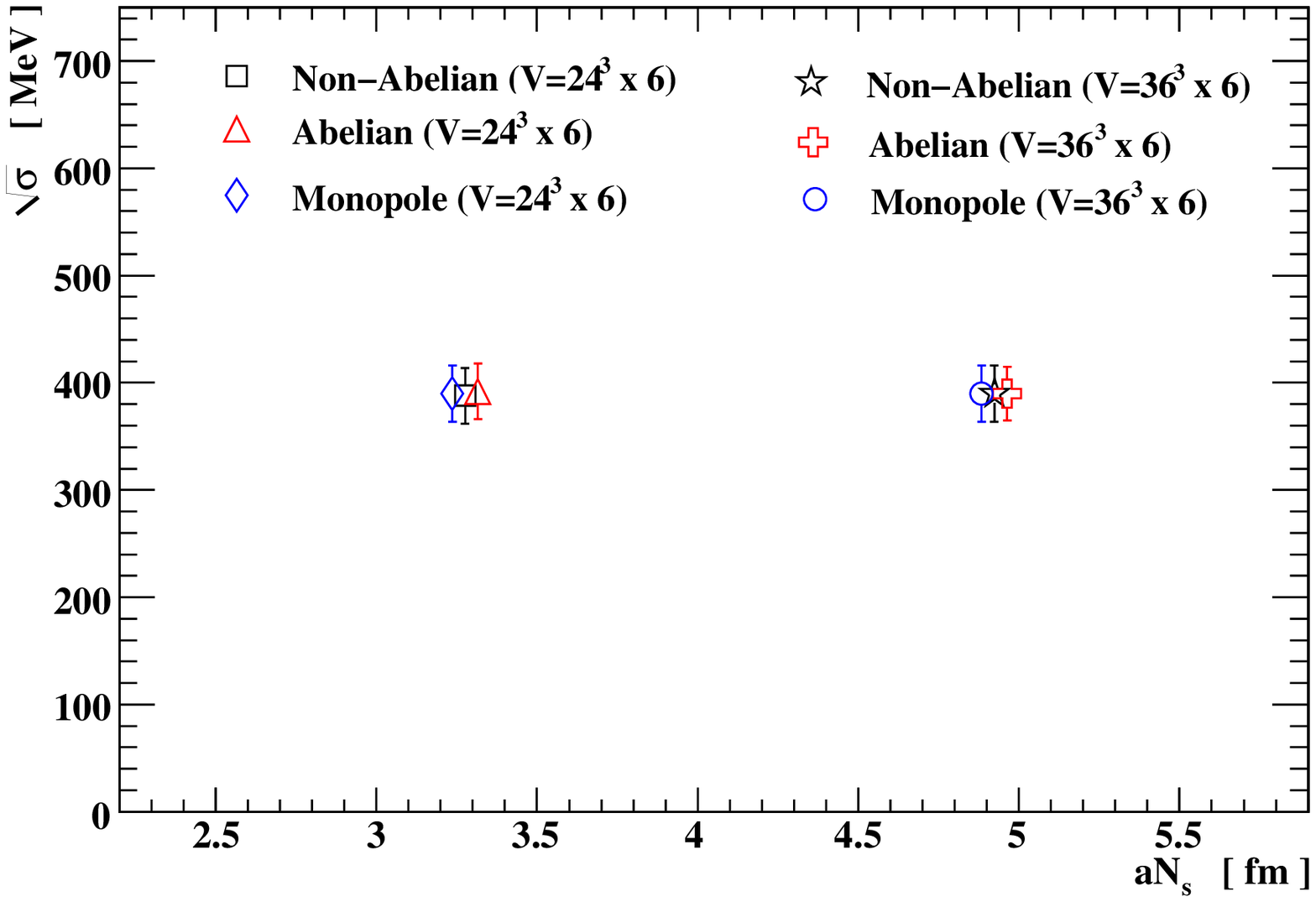}
\caption{\label{fig-6} The volume dependence of
the square root of the string tensions on the $24^{3}\times 6$ 
and $36^{3}\times 6$ lattices at $\beta=2.35$.}
\end{center}
\end{figure}

\par
Abelian dominance is seen again beautifully as in Sec.~\ref{sec:sec3}.
Moreover, we observe monopole dominance, i.e., 
the string tension of the static potential from the 
monopole Polyakov loop correlation function
is identical to that of the non-Abelian static potential,
while the potential from the photon Polyakov loop 
correlation function contains no linear part.
It is remarkable that Abelian dominance and monopole dominance
for the string tension are almost perfect as
explicitly shown in Fig.~\ref{fig-4}, which also 
show the good scaling behavior with respect to the change of lattice spacing.
We do not see the volume dependence of the string
tension as shown in Fig.~\ref{fig-6}.

\par
These results suggest that although the lattice monopoles defined 
in Eq.~\eqref{mon-current} are gauge-dependent, 
they contain physical gauge-invariant pieces responsible 
for confinement, which show up after taking the gauge average.

\section{The Abelian dual Meissner effect}
\label{sec:sec5}

\subsection{Correlation function for the field profile 
around the $q$-$\bar{q}$ system}

\begin{figure}[t]
\begin{center}  
\includegraphics[height=5.2cm]{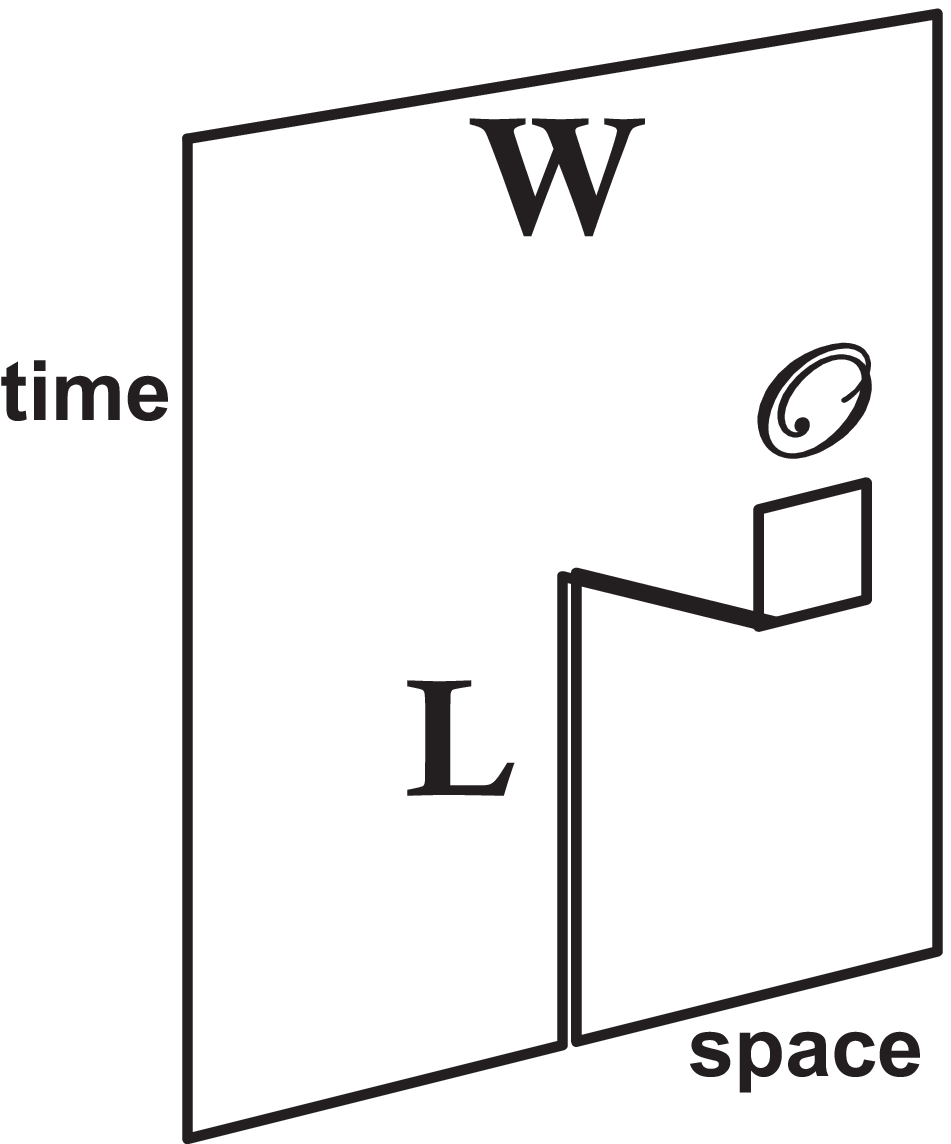}
\caption{\label{fig-7} A schematic figure for the connected correlation function.}
\end{center}
\begin{center}  
\includegraphics[height=5cm]{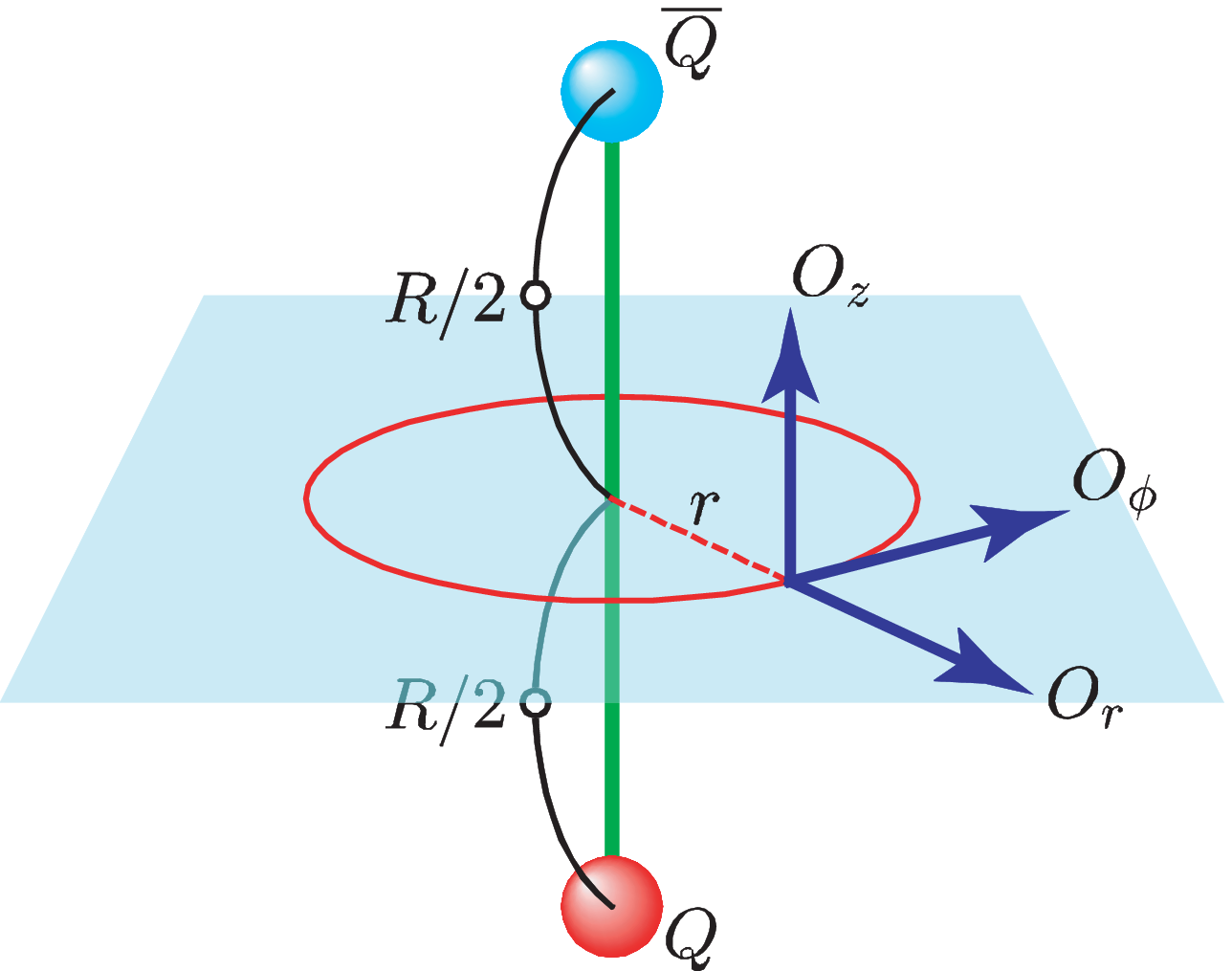}
\caption{\label{fig-8} Definition of the cylindrical 
coordinate $(r,\phi,z)$ along the $q$-$\bar{q}$ axis.}
\end{center}
\end{figure}

We investigate
the correlation function~\cite{Cea:1995zt, DiGiacomo:1989yp}
between a Wilson loop~$W$ and a local Abelian operator~${\cal O}$
connected by a product of non-Abelian link variables (Schwinger line)~$L$,
\begin{eqnarray}
\langle {\cal O}(r) \rangle_{W} 
= {\langle\mbox{Tr}\left[LW(R,T)L^{\dagger}\sigma^1{\cal O}(r)
\right]\rangle\over \langle 
\mbox{Tr}\left[W(R,T)\right]\rangle}\;.
\label{OW}
\end{eqnarray}
A schematic figure is depicted in Fig.~\ref{fig-7}.

\par
We shall use the cylindrical coordinate $(r,\phi,z)$ to parametrize 
the $q$-$\bar{q}$ system, where the $z$ axis
corresponds to the $q$-$\bar{q}$ axis and
$r$ to the transverse distance as shown in Fig.~\ref{fig-8}.
We are interested in the field profile as a function 
of $r$ on the mid-plane of the $q$-$\bar{q}$ system.

\subsection{Simulation parameters}

\begin{figure}[t]
\begin{center}   
\includegraphics[height=5.5cm]{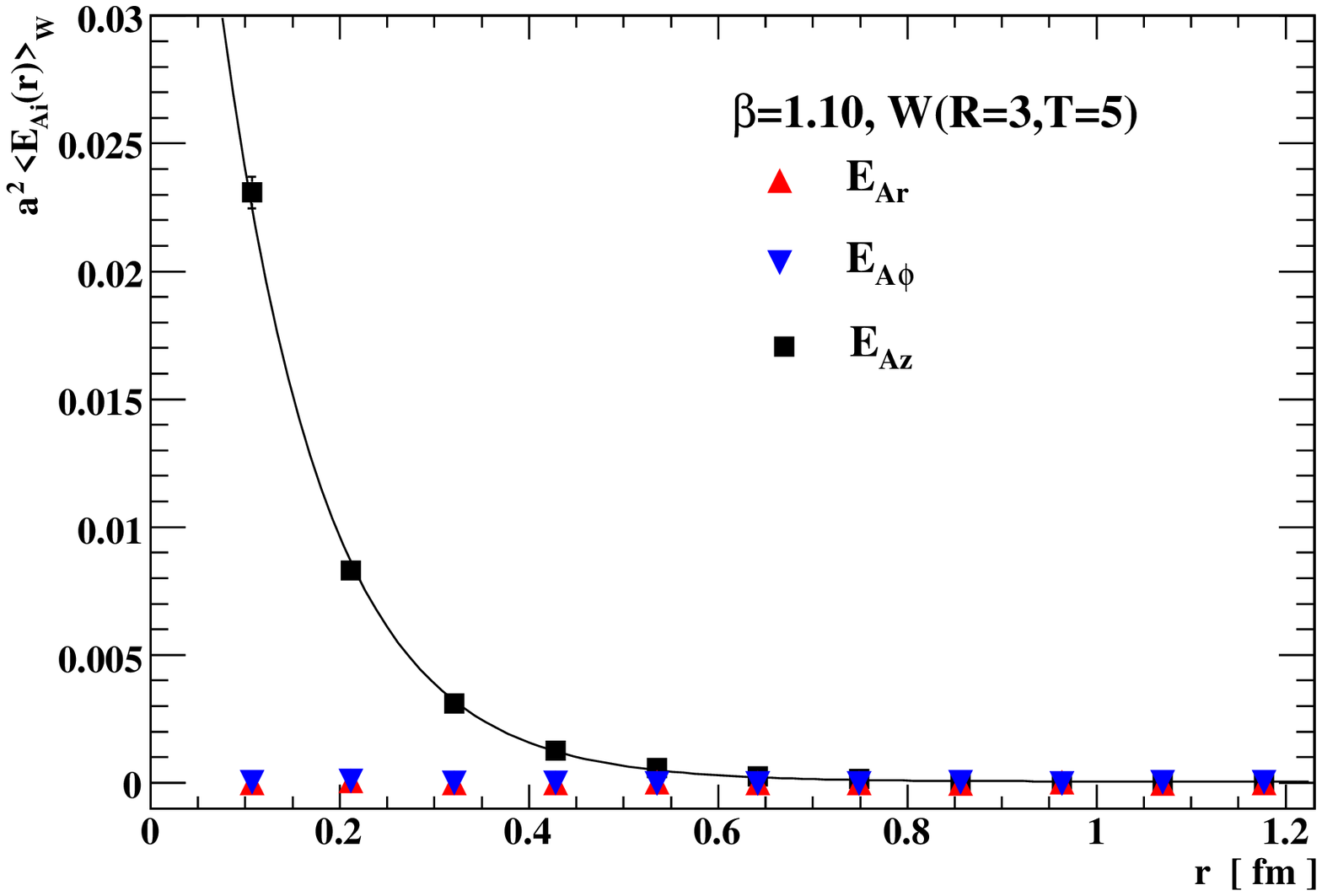}
\includegraphics[height=5.5cm]{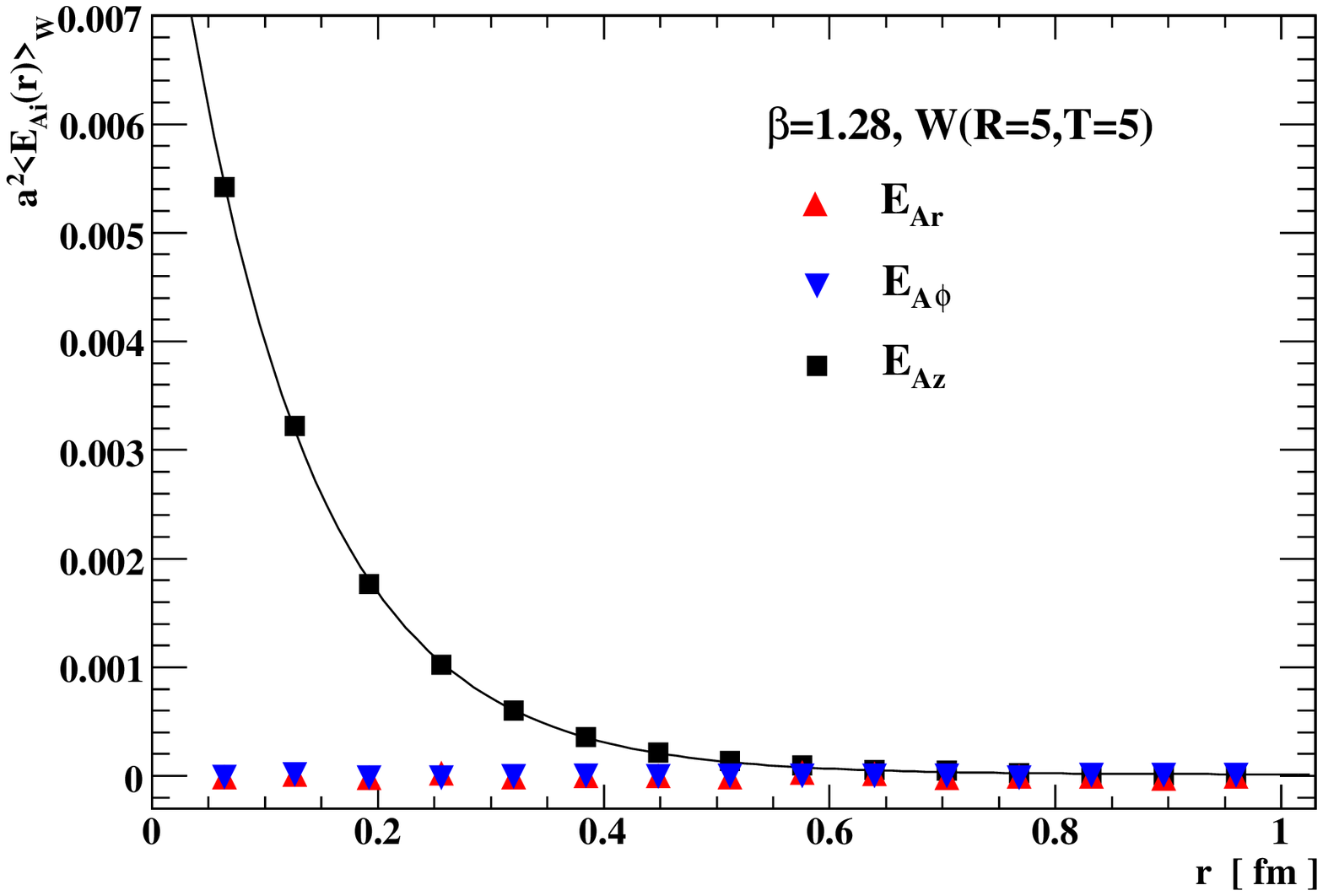}
\includegraphics[height=5.5cm]{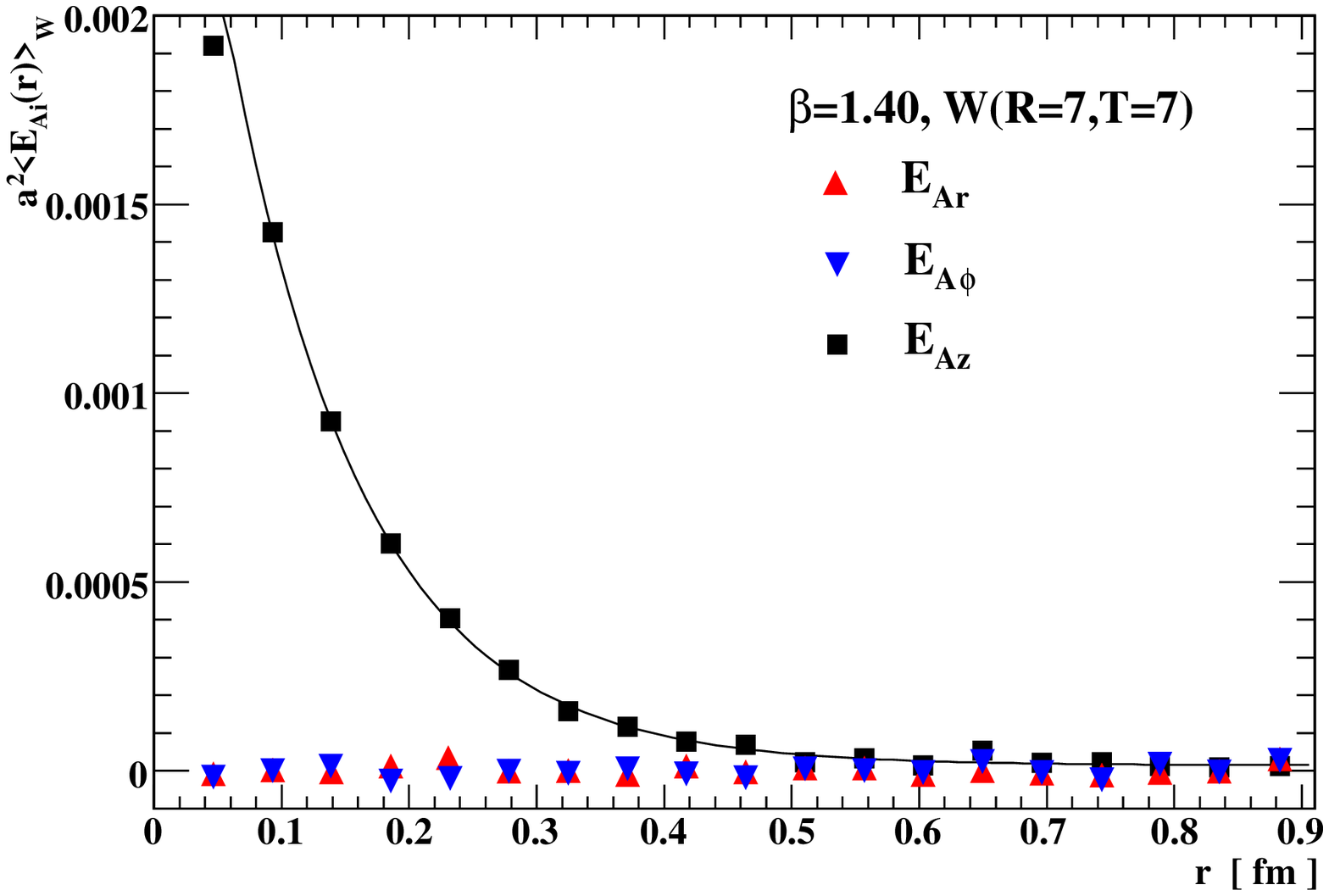}
\caption{\label{fig-9}
The profile of the color-electric field $\vec{E}_{\rm A}$ at 
$\beta=1.10$ (upper),  $\beta=1.28$ (middle) and $\beta=1.40$ (lower).}
\end{center}
\end{figure}

\par
In this computation, we employ the improved Iwasaki gauge
action~\cite{Iwasaki:1985we} with the coupling constants $\beta= 1.10$ and
$1.28$ on the $32^4$ lattice, and $\beta=1.40$ on the $40^4$ lattice
in order to investigate the scaling behavior of the correlation functions
with less finite lattice cutoff effects.
Simulation parameters are listed in Table~\ref{para-flux}.
The lattice spacings are determined so as to reproduce
the physical string tension \mbox{$\sqrt{\sigma}=440$~[MeV]}.
To improve the signal-to-noise ratio,
the APE smearing is applied to the spatial links 
of the Wilson loop~\cite{Albanese:1987ds}.
We use the Wilson loop $W(R=3,T=5)$ at $\beta=1.10$, $W(R=5,T=5)$ at $\beta=1.28$ 
and $W(R=7,T=7)$ at $\beta=1.40$.
Note that the physical $q$-$\bar{q}$ distance is the same ($R=0.32$~[fm])
for these Wilson loops.

\begin{table}[t]
\begin{center}
\caption{\label{para-flux}
Simulation parameters for the measurement of the field profile.
$n$ and $\alpha$ are the number of smearing steps and 
the smearing parameter, which are optimized to obtain reasonable signals.}
\begin{tabular}{c|c|c|c|c|c}
\hline
$\beta$ & $V$  & $a(\beta)$~[fm]& $N_{\rm conf}$ & $n$ & $\alpha$ \\
\hline
1.10 & $32^{4}$ & $0.1069(8)$& 5000 & 80 &0.2  \\
1.28 & $32^{4}$ & $0.0635(5)$ & 6000 & 80 &0.2  \\
1.40 & $40^{4}$ & $0.0465(2)$& 7996 & 80 &0.2  \\
\hline
\end{tabular}
\end{center}
\end{table}

\subsection{The penetration depth}

\par
\begin{figure}[t]
\begin{center}   
\includegraphics[height=5.8cm]{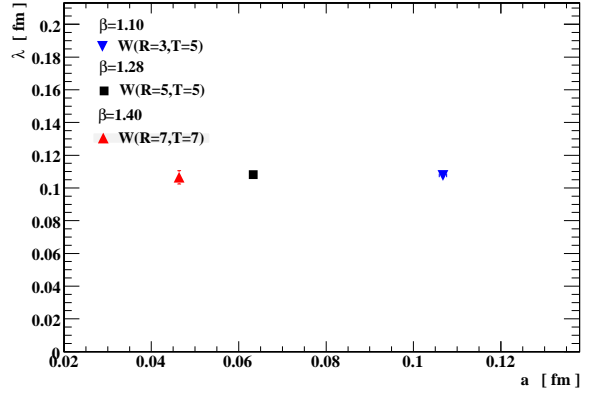}
\caption{\label{fig-10}
The penetration depth $\lambda$ as a function of lattice spacing $a(\beta)$.}
\end{center}
\end{figure}

We measure all cylindrical components of the color-electric fields 
${\cal O}(s)=E_{{\rm A}i}(s)=\bar{\Theta}_{4i}(s)$.
The results are plotted in Fig.~\ref{fig-9}.
We find that only $E_{{\rm A}z}$ has correlation with the Wilson loop.
We then fit $\langle E_{{\rm A}z} (r)\rangle_{W}$ to a function 
$f(r)=c_1\exp(-r/\lambda)+c_0$ and find that 
the profile of $\langle E_{{\rm A}z} (r)\rangle_{W}$
is well described by this functional form, i.e., 
the color-electric field is exponentially squeezed.
The fitting curves are also plotted in Fig.~\ref{fig-9}. 
The parameter $\lambda$ corresponds to the penetration depth and 
the values for three gauge couplings are summarized in Table~\ref{penetration}
and plotted in Fig.~\ref{fig-10} as a function of lattice spacing $a(\beta)$.
We find that the penetration depth~$\lambda$ shows the good scaling 
behavior.

\begin{table}[t]
\begin{center}
\caption{\label{penetration}
The parameter $\lambda$ corresponding to the penetration depth.} 
\begin{tabular}{c|c|c|c|c}
\hline
$\beta$& $W(R,T)$ &$\lambda$~[fm]& $c_1$ & $c_{0}$ \\ \hline
1.10 &$W(3,5)$ & 0.1075(13)& $6.09(18)\F\times 10^{-2}$ & $9(2)\F\times 10^{-5}$\\ 
1.28 &$W(5,5)$ & 0.1077(14)& $1.024(14)\times 10^{-2}$&$4.6(8)\times 10^{-6}$\\
1.40 &$W(7,7)$ & 0.106(4)\F\F & $3.40(17)\F\times 10^{-3}$& $1.6(8)\times 10^{-5}$\\ 
\hline
\end{tabular}
\end{center}
\end{table}

\subsection{The dual Amp\`ere law}
\par
\begin{figure}[t]
\begin{center}  
\includegraphics[height=5.5cm]{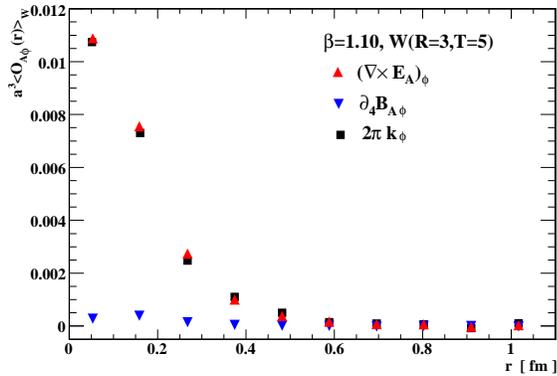}
\includegraphics[height=5.5cm]{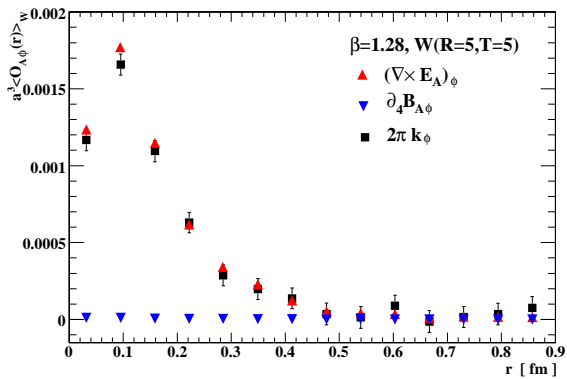}
\caption{\label{fig-11}
Tests of the dual Amp\`ere law at $\beta=1.10$ for $W(R=3,T=5)$ (upper)
and at $\beta=1.28$ for $W(R=5,T=5)$ (lower).}
\end{center}
\end{figure}

To see what squeezes the color-electric field, 
we study the Abelian (dual) Amp\`ere law derived
from the definition of the monopole current in Eq.~\eqref{mon-current},
\begin{eqnarray}
\vec{\nabla}\times\vec{E}_{\rm A}=
\partial_{4}\vec{B}_{\rm A}+2\pi\vec{k}\;, 
\end{eqnarray}
where $B_{{\rm A}i}(s)=(1/2)\epsilon_{ijk}\bar{\Theta}_{jk}(s)$.
The correlation of each term with the Wilson loop 
is evaluated on the same mid-plane of the $q$-$\bar{q}$ system 
as for the profile measurements of the color-electric field.
We find that only the azimuthal components are non-vanishing,
which are plotted in Fig.~\ref{fig-11}.
Note that if the color-electric field is purely of the Coulomb type, 
the curl of the electric field is zero. 
On the contrary, the curl of the electric field
is non-vanishing and is reproduced mostly by the monopole currents.
In any case, the dual Amp\`ere law is satisfied, which 
is a clear signal of the Abelian dual Meissner effect.
This result is quite the same as that 
observed in the MA gauge~\cite{Koma:2003gq,Koma:2003hv}. 

\subsection{The coherence length}

\begin{figure}[t]
\begin{center}  
\includegraphics[height=5.5cm]{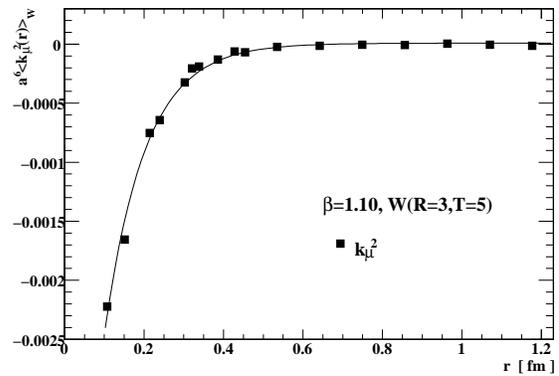}
\includegraphics[height=5.5cm]{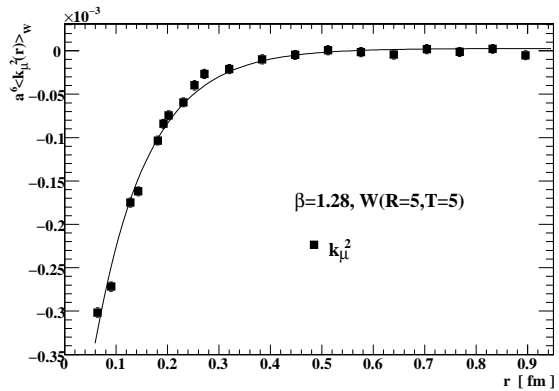}
\includegraphics[height=5.5cm]{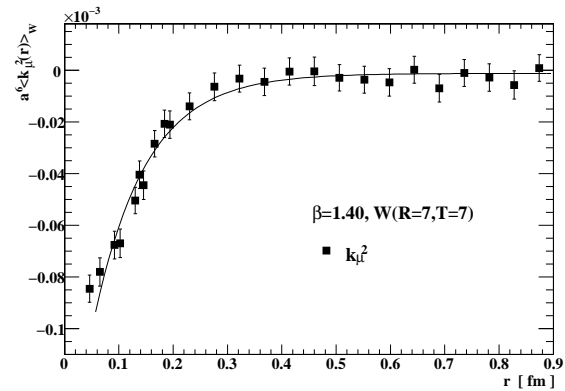}
\caption{\label{fig-13}
The profile of the squared monopole currents at $\beta=1.10$ (upper),
 $1.28$ (middle) and $1.40$ (lower).}
\end{center}
\end{figure}
Let us estimate the coherence length by evaluating 
the correlation function between
the squared monopole density~${\cal O}(s)=k_{\mu}^2(s)$
and the Wilson loop~\cite{Chernodub:2005gz}.
To measure such a correlation function, we use the disconnected
correlation function, since the Schwinger lines are
cancelled and the connected correlation functions for
the squared monopole currents are automatically reduced 
to the disconnected ones.
Simulation parameters, the lattice volume and the gauge couplings
are the same as the measurements of the color-electric field profile, 
but the number of gauge configurations is increased,
$N_{\rm conf}=5500$ for $\beta=1.10$ and 
$N_{\rm conf}=11887$ for $\beta=1.40$.
For $\beta=1.28$ we use the same number of 
configurations $N_{\rm conf}=6000$.
The physical $q$-$\bar{q}$ distance is again fixed 
to $R=0.32$~[fm].
To reduce the noise, we further produce $N_{\rm RGT}=100$
gauge copies for each independent configuration 
by applying the random gauge transformations and take
gauge-averaging.
\par
\begin{figure}[t]
\begin{center}  
\includegraphics[height=5.5cm]{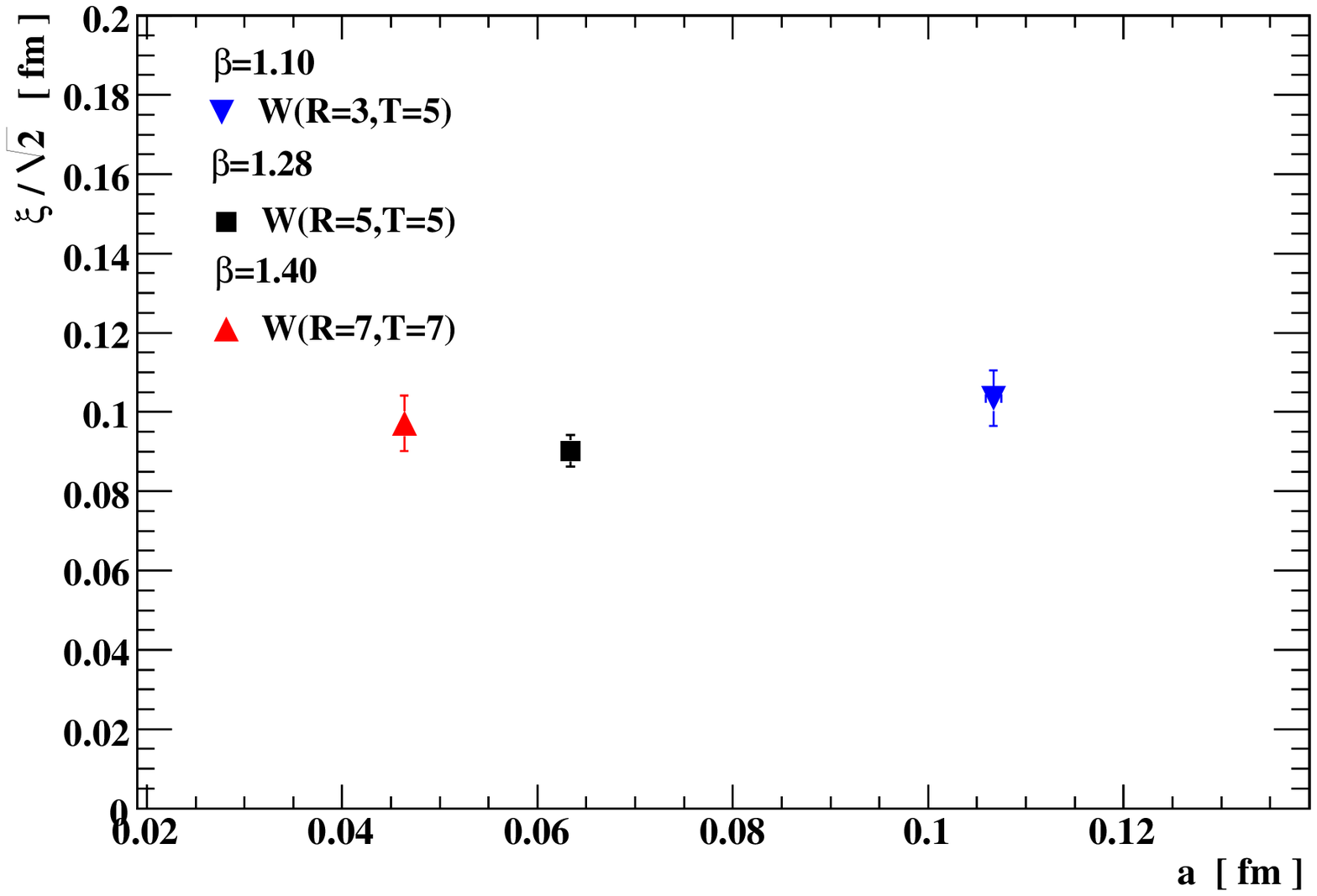}
\caption{\label{fig-14}The coherence length $\xi$
as a function of the lattice spacing $a(\beta)$.}
\end{center}
\begin{center}  
\includegraphics[height=5.6cm]{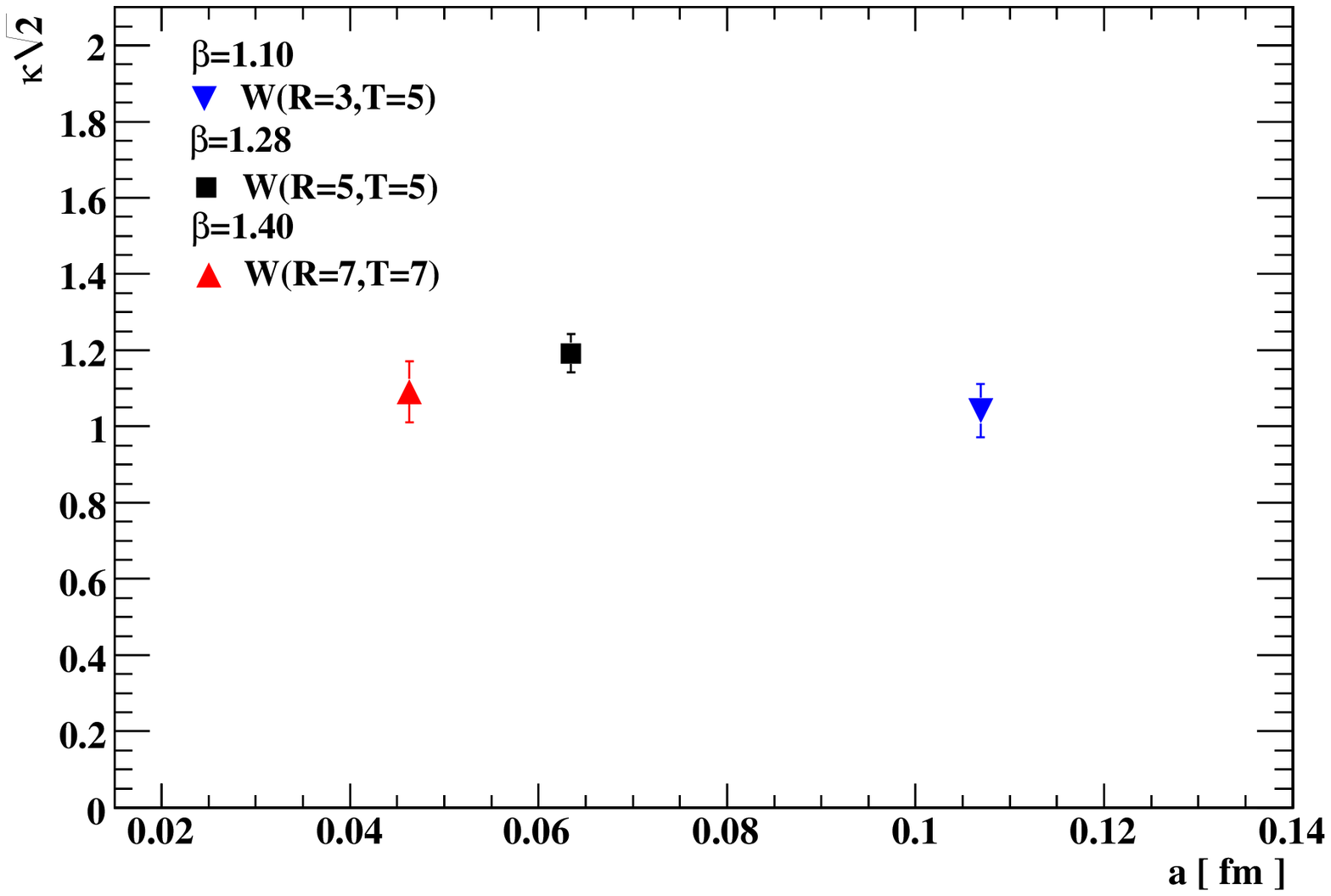}
\caption{\label{fig-15}
The GL parameters as a function of the lattice spacing $a(\beta)$.}
\end{center}
\end{figure}

The results are plotted in Fig.~\ref{fig-13}. 
We then fit the profile of $\langle k_{\mu}^2(r)\rangle_{W}$ to 
the functional form
$g(r)=c_{1}'\exp (-\sqrt{2}r/\xi) + c_{0}'$, where
the parameter $\xi$ corresponds to the coherence length.
We obtain the values for $\xi$ as summarized in Table~\ref{coherence_1}.
The coherence length shows the scaling behavior
as demonstrated in Fig.~\ref{fig-14} as a function of lattice 
spacing~$a(\beta)$.

\begin{table}[h]
\begin{center}
\caption{\label{coherence_1} 
The parameter $\xi/\sqrt{2}$
corresponding to the coherence length.}
\begin{tabular}{c|c|c|c|c}
\hline
$\beta$&$W(R,T)$ &$\xi/\sqrt{2}$~[fm] & $c_{1}'$ & $c_{0}'$ \\ 
\hline
1.10 &$W(3,5)$ & $0.103(7)$  & $-4.7(11)\times 10^{-3}$ 
& $-2(2)\times 10^{-6}$ \\ 
1.28 &$W(5,5)$ & $0.090(4)$    & $-7.5(3)\F\times 10^{-4}$ 
& ${\F}2(3)\times 10^{-6}$  \\ 
1.40 & $W(7,7)$ & $0.097(7)$ & $-1.68(16)\times 10^{-4}$ 
& $-1(3)\times 10^{-6}$\\ 
\hline
\end{tabular}
\end{center}
\end{table}

\subsection{The vacuum type}

\par
Taking the ratio of the penetration depth 
and the coherence length, the GL parameter 
$\sqrt{2}\kappa=\lambda/\xi$ can be estimated,
which characterizes the type of the superconducting vacuum.
The results are plotted in Fig.~\ref{fig-15} against lattice spacing $a(\beta)$.
We obtain $\sqrt{2}\kappa=1.04(7)$, $1.19(5)$ and $1.09(8)$
for $\beta=1.10$, $1.28$ and $1.40$, respectively.

\par
We find that the GL parameter shows the scaling behavior
and the value is about one.
This means that the vacuum type is near the border 
between the type~1 and~2 dual superconductor.
However, we note that the physical spatial size of the Wilson loop 
used in the present simulations is still small ($R=0.32$~[fm]).
Clearly, further quantitative studies with larger Wilson loops 
are needed to determine the definite value.

\section{Non-Abelian color confinement}\label{sec:sec6}

\par
Let us consider what is induced from the above numerical results.

\par
Since gauge fixing is not applied in these computations, 
Abelian fields in any color directions are equivalent.
Thus, our result is interpreted as that 
the color-electric fields in all color directions are squeezed 
and the Abelian (monopole) string tensions in all color directions 
are the same as the non-Abelian string tension. 
This indicates that QCD contains a gauge-invariant Abelian mechanism 
of confinement which is not related to the specific gauge fixing.
Namely Abelian monopoles in three color directions are condensed in the vacuum 
of the confinement phase of $SU(2)$ QCD.  

\par
Let us denote    quark fields having charge $1/2$ and $-1/2$ in the $\sigma_3$ direction, respectively, as $u_3$ and $d_3$. Then
local mesonic states, 
$u_3\bar{u}_3$ and $d_3\bar{d}_3$, are  Abelian color neutral
in the $\sigma_3$ direction. Consider next
\begin{eqnarray*}
u_1=\frac{u_3+d_3}{\sqrt{2}} \;, \quad d_1=\frac{u_3-d_3}{\sqrt{2}} \;,\\
u_2=\frac{iu_3+d_3}{\sqrt{2}} \;, \quad d_2=\frac{iu_3-d_3}{\sqrt{2}}.
\end{eqnarray*}
$u_1$ and $d_1$ ($u_2$ and $d_2$) are  quark fields having charge $1/2$ and $-1/2$ in the $\sigma_1$ ($\sigma_2/2$) direction. Using these expressions, the quark-gluon coupling term is written as
\begin{eqnarray}
\bar{\psi}\gamma^{\mu}\frac{\sigma^a}{2}\psi A^a_{\mu}&=&\frac{1}{2} 
(\bar{u}_3\gamma_{\mu}d_3+\bar{d}_3\gamma_{\mu}u_3)A^1_{\mu} \nonumber\\
&&
-i\frac{1}{2}(\bar{u}_3\gamma_{\mu}d_3-\bar{d}_3\gamma_{\mu}u_3)A^2_{\mu}\nonumber\\
&&+\frac{1}{2}(\bar{u}_3\gamma_{\mu}u_3-\bar{d}_3\gamma_{\mu}d_3)A^3_{\mu}\label{u3d3}\\
&=&\frac{1}{2}(\bar{u}_1\gamma_{\mu}u_1-\bar{d}_1\gamma_{\mu}d_1)A^1_{\mu}\nonumber\\
&&
+\frac{1}{2}(\bar{u}_2\gamma_{\mu}u_2-\bar{d}_2\gamma_{\mu}d_2)A^2_{\mu}\nonumber\\
&&+\frac{1}{2}(\bar{u}_3\gamma_{\mu}u_3-\bar{d}_3\gamma_{\mu}d_3)A^3_{\mu},
\end{eqnarray}  
where the first equation (\ref{u3d3}) is expressed in terms of $u_3$ and $d_3$ alone.
Consider local mesonic states $u_1\bar{u}_1$ and $d_1\bar{d}_1$ ($u_2\bar{u}_2$ and $d_2\bar{d}_2$) which
are Abelian color neutral in the $\sigma_1$ ($\sigma_2$) direction. When we look at the states
$u_1\bar{u}_1$ and $d_1\bar{d}_1$ in the   $\sigma_3$ direction, they 
are  written as the sum of  color-neutral and color-charged 
states:
\begin{eqnarray}
u_1\bar{u}_1 &=&
\frac{1}{2}(u_3\bar{u}_3 + d_3\bar{d}_3 + u_3\bar{d}_3 + d_3\bar{u}_3) \;, \\
d_1\bar{d}_1 &=&
\frac{1}{2}(u_3\bar{u}_3 + d_3\bar{d}_3 - u_3\bar{d}_3 - d_3\bar{u}_3)\; .
\end{eqnarray}
The same observation applies to the color-neutral states $u_2\bar{u}_2$ and $d_2\bar{d}_2$
in the $\sigma_2$ direction.
However, we find that
\begin{eqnarray}
u_1\bar{u}_1+d_1\bar{d}_1 =
u_2\bar{u}_2+d_2\bar{d}_2 =
u_3\bar{u}_3+d_3\bar{d}_3   \label{singlet}
\end{eqnarray}
are Abelian color neutral in all color directions.
The state (\ref{singlet}) is nothing but the non-Abelian color singlet state.

\par
This example tells us that
the Abelian color-neutral state in any color directions
corresponds to the physical non-Abelian color-singlet state.
Hence, the confinement of non-Abelian color charges can be 
explained in terms of the Abelian dual Meissner effect 
due to Abelian monopoles.
To the authors knowledge, this is the first paper that explains
the confinement of non-Abelian color charges only in terms 
of the Abelian dual Meissner effect.

\section{Concluding remarks }
\label{sec:sec7}

\par
We make some concluding remarks. 
The Abelian gauge fields extracted from the thermalized non-Abelian 
link fields contain originally topological monopoles 
responsible for the confinement mechanism of non-Abelian 
color charges even in the continuum limit.
Our results presented in this paper are almost the same
as those obtained in the maximally Abelian gauge. 
This suggests that the MA gauge fixing is the easiest method to
extract the physical ingredients of the monopoles, 
since we do not need very precise time-consuming
simulations in the MA gauge as done here. 

\par
In the lattice Landau gauge, it is known that no monopoles 
exist~\cite{Suzuki:2004dw} if one uses DeGrand-Toussaint definition
and the magnetic displacement current
takes a role of monopoles in the dual Amp\`ere law.
How to interpret the existence of a gauge-invariant Abelian confinement mechanism in the framework of the Landau gauge~?
Abelian monopoles are as a whole gauge-variant without gauge fixing, 
but they may contain a gauge-invariant physical component and 
a gauge-variant unphysical one.
The compatible interpretation would be that the Landau gauge is
a special gauge in which the unphysical gauge variant piece apparently 
cancels the physical one in the DeGrand-Toussaint monopoles,
but at the same time, 
the role of physical monopoles are carried by the color-magnetic displacement current,
which is just a matter of definition of monopoles on the lattice.
On the other hand,  in the MA gauge,  the main part of the
DeGrand-Toussaint monopoles is a physical component.

\par
If there exist physical gauge-invariant ingredients of Abelian monopoles, 
one could observe them in the real experiment~\cite{Chernodub:2008iv}.
To find the effect  in the confinement and also in the deconfinement phases 
is a very interesting topic in the future.

\begin{acknowledgments}
The numerical simulations of this work were done using RSCC computer
clusters in RIKEN and SX-8 computer at RCNP of Osaka University. 
The authors would like to thank RIKEN and RCNP for their
support of computer facilities.
The authors are also supported by JSPS and DFG under the Japan-Germany
Research Cooperative Program.
One of the authors (Y.K.) is partially supported by the 
Ministry of Education, Science, 
Sports and Culture, Japan, Grant-in-Aid 
for Young Scientists (B) (20740149).
\end{acknowledgments}


\begin{thebibliography}{10}

\bibitem{CMI:2000mp}
K.~Devlin,
\newblock \textit{The millennium problems : the seven greatest unsolved
  mathematical puzzles of our time},Basic Books, New York  (2002).

\bibitem{tHooft:1975pu}
G.~'t~Hooft,
\newblock in {\em Proceedings of the EPS International}, edited by A.~Zichichi,
  p. 1225, 1976.

\bibitem{Mandelstam:1974pi}
S.~Mandelstam,
\newblock Phys. Rept. {\bf 23}, 245 (1976).

\bibitem{'tHooft:1974qc}
G.~'t~Hooft,
\newblock Nucl. Phys. {\bf B79}, 276 (1974).

\bibitem{Polyakov:1976fu}
A.~M. Polyakov,
\newblock Nucl. Phys. {\bf B120}, 429 (1977).

\bibitem{Seiberg:1994rs}
N.~Seiberg and E.~Witten,
\newblock Nucl. Phys. {\bf B426}, 19 (1994).

\bibitem{tHooft:1981ht}
G.~'t~Hooft,
\newblock Nucl. Phys. {\bf B190}, 455 (1981).

\bibitem{Kronfeld:1987ri}
A.~S. Kronfeld, M.~L. Laursen, G.~Schierholz, and U.~J. Wiese,
\newblock Phys. Lett. {\bf B198}, 516 (1987).

\bibitem{Kronfeld:1987vd}
A.~S. Kronfeld, G.~Schierholz, and U.~J. Wiese,
\newblock Nucl. Phys. {\bf B293}, 461 (1987).

\bibitem{DeGrand:1980eq}
T.~A. DeGrand and D.~Toussaint,
\newblock Phys. Rev. {\bf D22}, 2478 (1980).

\bibitem{Suzuki:1992rw}
T.~Suzuki,
\newblock Nucl. Phys. Proc. Suppl. {\bf 30}, 176 (1993).

\bibitem{Singh:1993jj}
V.~Singh, D.~A. Browne, and R.~W. Haymaker,
\newblock Phys. Lett. {\bf B306}, 115 (1993).

\bibitem{Ejiri:1996sh}
S.~Ejiri, S.~Kitahara, T.~Suzuki, and K.~Yasuta,
\newblock Phys. Lett. {\bf B400}, 163 (1997), hep-lat/9608133.

\bibitem{Chernodub:1997ay}
M.~N. Chernodub and M.~I. Polikarpov,
\newblock (1997), hep-th/9710205.

\bibitem{Bali:1997cp}
G.~S. Bali, C.~Schlichter, and K.~Schilling,
\newblock Prog. Theor. Phys. Suppl. {\bf 131}, 645 (1998).

\bibitem{Suzuki:1998hc}
T.~Suzuki,
\newblock Prog. Theor. Phys. Suppl. {\bf 131}, 633 (1998).

\bibitem{Koma:2003gq}
Y.~Koma, M.~Koma, E.-M. Ilgenfritz, T.~Suzuki, and M.~I. Polikarpov,
\newblock Phys. Rev. {\bf D68}, 094018 (2003), 

\bibitem{Koma:2003hv}
Y.~Koma, M.~Koma, E.-M. Ilgenfritz, and T.~Suzuki,
\newblock Phys. Rev. {\bf D68}, 114504 (2003), 

\bibitem{Kitahara:1994vt}
S.~Kitahara, Y.~Matsubara, and T.~Suzuki,
\newblock Prog. Theor. Phys. {\bf 93}, 1 (1995), hep-lat/9411036.

\bibitem{Stack:1994wm}
J.~D. Stack, S.~D. Neiman, and R.~J. Wensley,
\newblock Phys. Rev. {\bf D50}, 3399 (1994), hep-lat/9404014.

\bibitem{Shiba:1994ab}
H.~Shiba and T.~Suzuki,
\newblock Phys. Lett. {\bf B333}, 461 (1994), hep-lat/9404015.

\bibitem{Shiba:1994db}
H.~Shiba and T.~Suzuki,
\newblock Phys. Lett. {\bf B351}, 519 (1995), hep-lat/9408004.

\bibitem{Suzuki:1994ay}
T.~Suzuki, S.~Ilyar, Y.~Matsubara, T.~Okude, and K.~Yotsuji,
\newblock Phys. Lett. {\bf B347}, 375 (1995).

\bibitem{Ejiri:1994uw}
S.~Ejiri, S.~Kitahara, Y.~Matsubara, and T.~Suzuki,
\newblock Phys. Lett. {\bf B343}, 304 (1995), hep-lat/9407022.

\bibitem{Bali:1996mv}
G.~S. Bali and C.~Schlichter,
\newblock Prog. Theor. Phys. Suppl. {\bf 122}, 67 (1996).

\bibitem{Kato:1998ur}
S.~Kato, N.~Nakamura, T.~Suzuki, and S.~Kitahara,
\newblock Nucl. Phys. {\bf B520}, 323 (1998).

\bibitem{Chernodub:2000ax}
M.~N. Chernodub {\em et~al.},
\newblock Phys. Rev. {\bf D62}, 094506 (2000).

\bibitem{Ishiguro:2001jd}
K.~Ishiguro, T.~Suzuki, and T.~Yazawa,
\newblock JHEP {\bf 01}, 038 (2002), hep-lat/0112022.

\bibitem{Ezawa:1982bf}
Z.~F. Ezawa and A.~Iwazaki,
\newblock Phys. Rev. {\bf D25}, 2681 (1982).

\bibitem{Suzuki:1988yq}
T.~Suzuki,
\newblock Prog. Theor. Phys. {\bf 80}, 929 (1988).

\bibitem{Maedan:1988yi}
S.~Maedan and T.~Suzuki,
\newblock Prog. Theor. Phys. {\bf 81}, 229 (1989).

\bibitem{Chernodub:2003mm}
M.~N. Chernodub,
\newblock Phys. Rev. {\bf D69}, 094504 (2004), hep-lat/0308031.

\bibitem{Sekido:2007mp}
T.~Sekido, K.~Ishiguro, Y.~Koma, Y.~Mori, and T.~Suzuki,
\newblock Phys. Rev. {\bf D76}, 031501 (2007).

\bibitem{Carmona:2001ja}
J.~M. Carmona, M.~D'Elia, A.~Di~Giacomo, B.~Lucini, and G.~Paffuti,
\newblock Phys. Rev. {\bf D64}, 114507 (2001).

\bibitem{Cea:2000zr}
P.~Cea and L.~Cosmai,
\newblock Phys. Rev. {\bf D62}, 094510 (2000).

\bibitem{Suzuki:2007jp}
T.~Suzuki, K.~Ishiguro, Y.~Koma, and T.~Sekido,
\newblock Phys. Rev. {\bf D77}, 034502 (2008), 0706.4366.

\bibitem{Luscher:2001up}
M.~L$\ddot{\mbox{u}}$scher and P.~Weisz,
\newblock JHEP {\bf 09}, 010 (2001).

\bibitem{Necco:2001xg}
S.~Necco and R.~Sommer,
\newblock Nucl. Phys. {\bf B622}, 328 (2002).

\bibitem{Luscher:2002qv}
M.~L$\ddot{\mbox{u}}$scher and P.~Weisz,
\newblock JHEP {\bf 07}, 049 (2002).

\bibitem{Ogilvie:1998wu}
M.~C. Ogilvie,
\newblock Phys. Rev. {\bf D59}, 074505 (1999).

\bibitem{Faber:1998en}
M.~Faber, J.~Greensite, and S.~Olejnik,
\newblock JHEP {\bf 01}, 008 (1999).

\bibitem{Hasenfratz:2001hp}
A.~Hasenfratz and F.~Knechtli,
\newblock Phys. Rev. {\bf D64}, 034504 (2001).

\bibitem{Cea:1995zt}
P.~Cea and L.~Cosmai,
\newblock Phys. Rev. {\bf D52}, 5152 (1995).

\bibitem{DiGiacomo:1989yp}
A.~Di~Giacomo, M.~Maggiore, and S.~Olejnik,
\newblock Phys. Lett. {\bf B236}, 199 (1990).

\bibitem{Iwasaki:1985we}
Y.~Iwasaki,
\newblock Nucl. Phys. {\bf B258}, 141 (1985).

\bibitem{Albanese:1987ds}
APE, M.~Albanese {\em et~al.},
\newblock Phys. Lett. {\bf B192}, 163 (1987).

\bibitem{Chernodub:2005gz}
M.~N. Chernodub {\em et~al.},
\newblock Phys. Rev. {\bf D72}, 074505 (2005).

\bibitem{Suzuki:2004dw}
T.~Suzuki, K.~Ishiguro, Y.~Mori, and T.~Sekido,
\newblock Phys. Rev. Lett. {\bf 94}, 132001 (2005), 

\bibitem{Chernodub:2008iv}
M.~N. Chernodub, A.~Nakamura, and V.~I. Zakharov,
\newblock Phys. Rev. {\bf D78}, 074021 (2008), 0807.5012.

\end{thebibliography}

\end{document}